\documentclass[twocolumn]{aastex63}

\usepackage[fleqn]{amsmath}
\usepackage[hyphenbreaks]{breakurl}
\usepackage{float}

\makeatletter
\newcommand*{\rom}[1]{\expandafter\@slowromancap\romannumeral #1@}
\makeatother

\received{Dec 23, 2020}
\revised{Jan 19, 2021}
\accepted{Jan 19, 2021}

\shorttitle{$JHK_s$-band PLRs for RR Lyrae stars in M53}
\shortauthors{Bhardwaj A. et al.}

\begin{document}
\title{RR Lyrae variables in Messier 53: Near-infrared Period--Luminosity relations and the calibration using {\it Gaia} Early Data Release 3}

\correspondingauthor{Anupam Bhardwaj}
\email{anupam.bhardwajj@gmail.com; abhardwaj@kasi.re.kr}
\author[0000-0001-6147-3360]{Anupam Bhardwaj}\thanks{IAU Gruber Foundation Fellow}\thanks{EACOA Fellow}
\affil{Korea Astronomy and Space Science Institute, Daedeokdae-ro 776, Yuseong-gu, Daejeon 34055, Republic of Korea}
\author[0000-0002-6577-2787]{Marina Rejkuba}
\affiliation{European Southern Observatory, Karl-Schwarzschild-Stra\ss e 2, 85748, Garching, Germany}
\author[0000-0002-7203-5996]{Richard de Grijs}
\affiliation{Department of Physics and Astronomy, Macquarie University, Balaclava Road, Sydney, NSW 2109, Australia}
\affiliation{Research Centre for Astronomy, Astrophysics and Astrophotonics, Macquarie University, Balaclava Road, Sydney, NSW 2109, Australia}
\affiliation{International Space Science Institute--Beijing, 1 Nanertiao, Zhongguancun, Hai Dian District, Beijing 100190, China}
\author{Soung-Chul Yang}
\affiliation{Korea Astronomy and Space Science Institute, Daedeokdae-ro 776, Yuseong-gu, Daejeon 34055, Republic of Korea}
\author[0000-0001-6147-3360]{Gregory J. Herczeg}
\affil{Kavli Institute for Astronomy and Astrophysics, Peking University, Yi He Yuan Lu 5, Hai Dian District, Beijing 100871, China}
\author[0000-0002-1330-2927]{Marcella Marconi}
\affil{INAF-Osservatorio astronomico di Capodimonte, Via Moiariello 16, 80131 Napoli, Italy}
\author[0000-0001-6802-6539]{Harinder P. Singh}
\affiliation{Department of Physics and Astrophysics, University of Delhi,  Delhi-110007, India }
\author{Shashi Kanbur}
\affiliation{Department of Physics, State University of New York, Oswego, NY 13126, USA}
\author[0000-0001-8771-7554]{Chow-Choong Ngeow}
\affil{Graduate Institute of Astronomy, National Central University, 300 Jhongda Road, 32001 Jhongli, Taiwan}

\begin{abstract} 

We present new near-infrared, $JHK_s$, Period--Luminosity relations (PLRs) for RR Lyrae variables in the Messier 53 (M53 or NGC 5024) globular cluster. Multi-epoch $JHK_s$ observations, obtained with the WIRCam instrument on the 3.6-m Canada France Hawaii Telescope, are used for the first time to estimate precise mean-magnitudes for 63 RR Lyrae stars in M53 including 29 fundamental-mode (RRab) and 34 first-overtone mode (RRc) variables. The $JHK_s$-band PLRs for RR Lyrae stars are best constrained for RRab types with a minimal scatter of 22, 23, and 19 mmag, respectively. The combined sample of RR Lyrae is used to derive the $K_s$-band PLR, $K_s = -2.303 (0.063) \log P + 15.212 (0.016)$ exhibiting a $1\sigma$ dispersion of only $0.027$ mag. Theoretical Period--Luminosity--Metallicity (PLZ) relations are used to predict parallaxes for 400 Galactic RR Lyrae resulting in a median parallax zero-point offset of $-7\pm3~\mu$as in {\it Gaia} Early Data Release 3 (EDR3), which increases to $22\pm2~\mu$as if the parallax corrections are applied. We also estimate a robust distance modulus, $\mu_\textrm{M53} = 16.403 \pm 0.024$ (statistical) $\pm 0.033$ (systematic) mag, to M53 based on theoretical calibrations. Homogeneous and precise mean-magnitudes for RR Lyrae in M53 together with similar literature data for M3, M4, M5 and $\omega$ Cen are used to empirically calibrate a new RR Lyrae PLZ$_{K_s}$ relation, $K_s = -0.848 (0.007) -2.320 (0.006) \log P + 0.166 (0.011) {\rm[Fe/H]}$, anchored with {\it Gaia} EDR3 distances and theoretically predicted relations, and simultaneously estimate precise RR Lyrae based distances to these globular clusters.\\

\end{abstract}

\section{Introduction}

Classical pulsating stars such as RR Lyrae variables are excellent distance indicators thanks to their well-defined Period--Luminosity relations (PLRs) especially at near-infrared (NIR) wavelengths \citep[see reviews by][]{beaton2018, bhardwaj2020}. NIR photometry offers several advantages, firstly due to lower sensitivity to reddening than at optical wavelengths resulting in smaller uncertainties owing to extinction. Secondly, theoretical and empirical PLRs of RR Lyrae display a steady decrease in the dispersion moving from optical to infrared wavelengths. This suggests that the temperature or color variations due to the finite width of the instability strip are smaller at infrared wavelengths, thus, supporting the basic assumption of negligible luminosity variation in the PLRs as a function of temperature for a given period. Furthermore, NIR light curves of RR Lyrae are more sinusoidal exhibiting small amplitude variations ($Amp_{K_s} < 0.4$~mag) making it is easier to determine accurate mean-magnitudes with fewer observations. Thanks to these advantages and the increase in the volume of NIR observations, RR Lyrae have gained a significant boost as distance indicators in the past two decades \citep{sollima2006, coppola2011, braga2015, muraveva2015, navarrete2017, braga2018, muraveva2018a, bhardwaj2020a}. 

RR Lyrae are typically the most numerous variable stars in Globular Clusters (GCs). GCs play a crucial role in our understanding of, for example, stellar evolution \citep{dotter2010, denissenkov2017}, multiple stellar populations \citep{gratton2012, bastian2017}, and Galaxy formation and evolution \citep{massari2019, kruijssen2019}. Given that GCs are of great interest for a wide range of astrophysical studies, several detailed investigations have been carried out regarding their ages, metallicities, and structure and kinematics \citep[e.g.,][]{sarajedini2007, carretta2009, marin2009, vandenberg2013, helmi2018}. However, homogeneous and accurate distances to GCs are limited to nearby systems \citep{helmi2018}. RR Lyrae variables, being excellent standard candles, can be used to obtain precise and accurate distances to GCs \citep[e.g.,][]{braga2015} provided the absolute calibration of their Period--Luminosity--Metallicity (PLZ) relations in the Mily Way. {\it Gaia} parallaxes from current \citep{brown2020, lindegren2020} and future data releases will play a critical role in the calibration of RR Lyrae PLZ relations. Absolute distances to GCs are important to constrain their intrinsic properties such as their absolute ages which also place a lower limit on the age of the Universe \citep{krauss2003, marin2009}.

Messier 53 (M53 or NGC 5024) is a very old \citep[$\sim 12.25\pm 0.25$ Gyr,][]{vandenberg2013} and metal-poor \citep[{[Fe/H]}$\sim-2.06$~dex,][]{harris2010, boberg2016} GC which hosts more than 60 RR Lyrae variables \citep[][]{ferro2011}. M53 is favorably located in the Galactic halo at a very high latitude, relatively far from the Galactic plane, therefore its interstellar reddening is small, $E(B-V)\sim 0.02$~mag \citep{harris2010}, and the contamination from field stars is expected to be minimal. Several optical photometric studies have targeted the relatively large variable population of M53 \citep{kopacki2000, dekany2009, safonova2011, ferro2011, bramich2012}, investigating, for example, optical Period--Luminosity--Color (PLC) relations and Blazhko variations among RR Lyare stars. M53 hosts one of the largest samples of Blazhko RRc stars \citep{ferro2012}. However, variable stars in M53 have not been explored at NIR wavelengths to date. 
  
In this paper, we present NIR pulsation properties and the PLRs of RR Lyrae variables in M53 for the first time. New NIR PLRs of M53 RR Lyrae provide a robust estimate of the true distance modulus to this cluster and a test of the universality of PLRs when compared with similar literature data in other GCs. We also provide a new empirical calibration of PLZ relation in the $K_s$-band for RR Lyrae in the GCs using {\it Gaia} Early Data Release 3 (EDR3) parallaxes. The paper is organized as follows. In Section~\ref{sec:data}, we discuss the NIR photometry and pulsation properties of RR Lyrae stars. The $JHK_s$-band PLRs for M53 RR Lyrae and a robust distance estimate to the cluster are presented in Section~\ref{sec:RR Lyrae_plr}. The calibration of PLZ relations for RR Lyrae after combining M53 photometry with literature data is discussed in Section~\ref{sec:plzr}. The main results of this work are summarized in Section~\ref{sec:discuss}.

\section{RR Lyrae Photometry} \label{sec:data}

We used NIR time-series observations of RR Lyrae in M53 obtained using the WIRCam instrument \citep{puget2004} mounted on the 3.6-m Canada France Hawaii Telescope (CFHT). Our observations were 
collected from 26 to 29 May 2019 during the same observing run and following the same observing strategy as for Messier 3 described in \citet{bhardwaj2020a}. In brief, we covered $\sim21'\times 21'$ around the cluster center and obtained on average 20 epochs in the $JHK_s$-bands. Point-spread function (PSF) photometry was performed on the images using {\texttt{DAOPHOT/ALLSTAR}} and {\texttt{ALLFRAME}} software \citep{stetson1987, stetson1994}. The details regarding the photometric data reduction, the PSF photometry, and the photometric calibrations in the 2MASS system are similar to those pertaining to M3 and are discussed in detail in \citet{bhardwaj2020a}. 

\begin{figure}
\epsscale{1.2}
\plotone{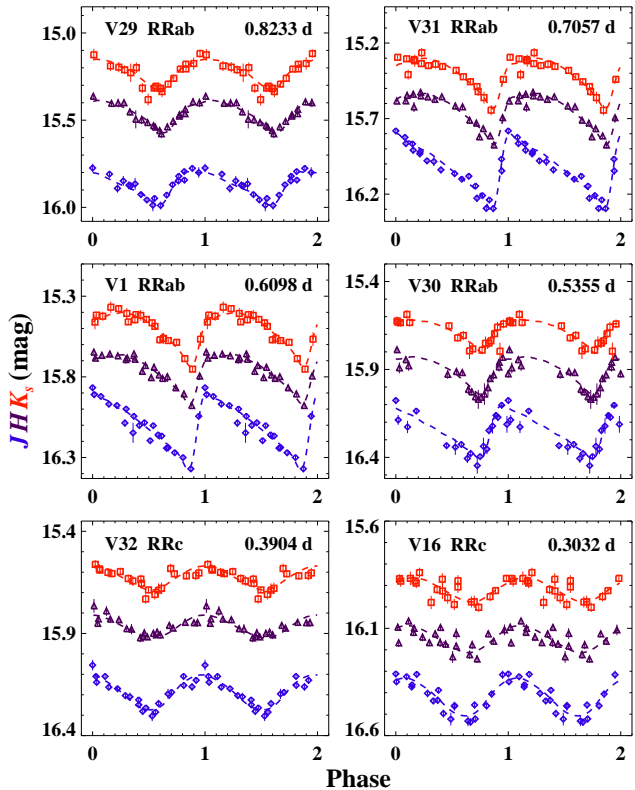}
\caption{Example $JHK_s$-band light curves of M53 RR Lyrae covering the entire range of periods in our sample. The $J$ (blue) and $K_s$ (red) light curves are offset for clarity by $+0.1$ and $-0.2$ mag, respectively. The dashed lines represent the best-fitting templates to the data in each band. Star ID, subtype, and the pulsation period are included at the top of each panel.} 
\label{fig:lcs}
\end{figure}

\subsection{RR Lyrae sample and the light curves}

Our sample of RR Lyrae candidates in M53 was adopted from the catalog{\footnote{\url{http://www.astro.utoronto.ca/~cclement/}}} of \citet{clement2001} which was last updated in 2012 for M53 variables 
following the work of \citet{ferro2012}. There are 64 RR Lyrae (29 RRab and 35 RRc) in M53 for which coordinates and periods are available. One RRc (V48) is outside our field of view and thus our final sample contains 63 RR Lyrae stars. NIR light curves of these variables were extracted using a cross-match with the photometric catalog within a tolerance of $1\arcsec$. The majority of RR Lyrae (58 of 63) match within a tolerance of $0.1\arcsec$ while one object (V53), located in the central crowded region, was retrieved with $\Delta \sim 1.6\arcsec$.  

RR Lyrae light curves were phased using periods adopted from \citet{clement2001} with the reference epoch corresponding to our first observation. No distinct periodic variability is seen for variables (V72, V91, V92) which have small optical amplitudes ($Amp_V \lesssim 0.2$~mag). \citet{ferro2011} provided a detailed description of the photometric contamination for a number of RR Lyrae stars - V60, V61, V62, V64, V70, V71, due to blending with nearby stars. In our photometry, V53, V61, and V64 do not show any distinct variability but periodicity is obtained for V60, V62, and V71 albeit with large scatter. Since M53 RR Lyrae do not have periods close to 0.5 days, for example like M3, light curves are well-sampled with no large phase gaps especially around maximum and minimum. Light curves of 57 RR Lyrae with full phase coverage were assigned a quality flag `A' while remaining 6 stars exhibiting significantly contaminated photometric light curves were flagged as `B'. 

NIR templates from \citet{braga2019} were fitted to the phased light curves of RR Lyrae solving for amplitude and phase simultaneously. There is strong evidence that the amplitude ratios or the light
curve shapes of RRab change at different periods in Oosterhoff I (OoI) and II (OoII) type clusters \citep[for example, M3 and $\omega$ Cen in][]{braga2018, bhardwaj2020a}. Therefore, we fit all three templates for RRab from \citet{braga2019} to our light curves irrespective of their periods and adopt the best-fitting template. Fig.~\ref{fig:lcs} shows a few example light curves of different RR Lyrae subclasses spanning the entire period range ($0.27<P<0.83$~days). For light curves with quality-flag `A', best-fitting templates were adopted for all RR Lyrae to determine mean magnitudes and amplitudes. No amplitude measurements are provided for light curves with flag `B', and only the error weighted mean is considered for further analysis. We also used the epoch of maximum-light in the $V$-band from \citet{ferro2012} to constrain the phase parameter, thus solving only for variable amplitudes in the NIR bands. However, no statistically significant difference was found in the estimated $JHK_s$-band mean-magnitudes of RR Lyrae variables. Median photometric uncertainties were added to the errors on the mean magnitudes resulting from the best-fitting templates. The NIR pulsation properties of RR Lyrae in M53 are tabulated in Table~\ref{tbl:m53_nir}. Time-series NIR data of known SX Phe \citep{ferro2011} and other candidate variables will be presented in a future study together with long-term optical photometry.

\subsection{Color--magnitude and Bailey diagrams}

\begin{figure}
\epsscale{1.2}
\plotone{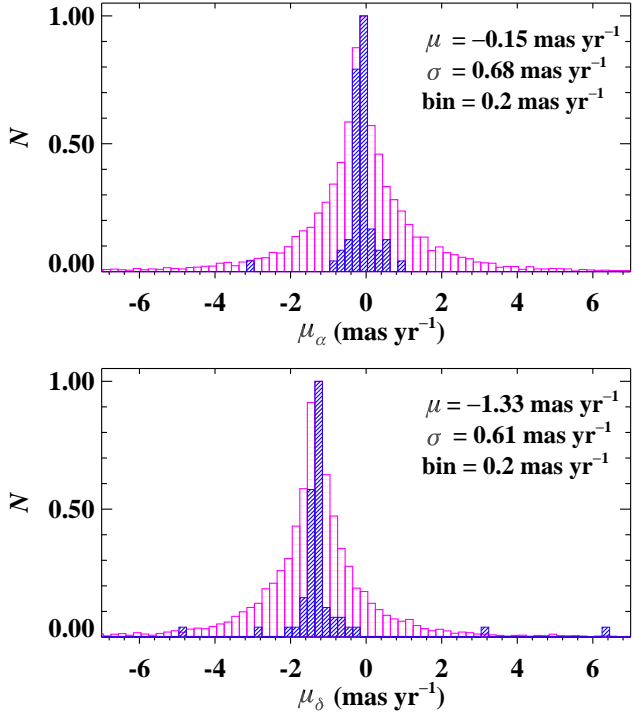}
\caption{Normalized histograms of proper motions along right ascension ({top}) and declination ({bottom}) for all M53 sources in the {\it Gaia} EDR3 within the WIRCam field of view. The mean values and the standard deviations of the Gaussian fits to the histograms are also shown in each panel. The histograms of proper motions of RR Lyrae variables in the M53 cluster are also overplotted as slanted blue lines.}
\label{fig:pms}
\end{figure}

Our photometric catalog was cross-matched with the latest {\it Gaia} EDR3 \citep[][]{lindegren2020} to obtain proper motions for sources in the field of view of M53. Fig.~\ref{fig:pms} shows the normalized histograms of proper motions for M53 sources. The mean proper motions along the right ascension and declination axes for M53 sources were $\mu_\alpha = -0.154\pm0.007$ and $\mu_\delta = -1.333\pm0.006$ mas yr$^{-1}$ with a Gaussian half width at half maximum of $0.675$ and $0.613$~mas yr$^{-1}$, respectively. These mean proper motion estimates for M53 are in excellent agreement with those from \citet[][ $\mu_\alpha = -0.147\pm0.005$ and $\mu_\delta = -1.351\pm0.003$ mas yr$^{-1}$]{helmi2018} using {\it Gaia} DR2. However, the scatter in the mean proper motions is reduced significantly with EDR3 data and therefore sources within the 5 and 95 percentiles were considered as the members of the M53 cluster. The histograms of proper motions of RR Lyrae variables in the M53 cluster are also overplotted and only V54 is a proper motion outlier along both the right ascension and declination axes. Furthermore, V60 and V64 also have proper motions beyond $3\sigma$ of the mean value along the declination axis.

\startlongtable
\begin{deluxetable*}{rccccccccccccccr}
\tablecaption{NIR pulsation properties of RR Lyrae in the M53 cluster. \label{tbl:m53_nir}}
\tabletypesize{\footnotesize}
\tablewidth{0pt}
\tablehead{\colhead{ID} & \colhead{RA} & \colhead{Dec} & \colhead{Period} & \colhead{Type}& \multicolumn{3}{c}{Mean magnitudes}  & \multicolumn{3}{c}{$\sigma_{\textrm{mag}}$} & \multicolumn{3}{c}{Amplitudes ($Amp_\lambda$)}  & $\Delta''$ & \colhead{QF}\\
 	&	&		& 	    &	   & $J$    & $H$     & $K_s$ & $J$     & $H$    & $K_s$ & $J$     & $H$    & $K_s$  &        &	\\  
 	&	deg.&	deg.	& days 	    &	   & \multicolumn{3}{c}{mag}  & \multicolumn{3}{c}{mag}  & \multicolumn{3}{c}{mag}   &  arcsec      &  }
\startdata
   V1 &  198.234667 &   18.120528 & 0.60980 &  RRab & 16.006 & 15.765 & 15.710 &  0.014 &  0.016 &  0.022 &  0.480 &  0.305 &  0.338 &  0.004 &   A\\
   V2 &  198.209500 &   18.116917 & 0.38620 &   RRc & 16.119 & 15.909 & 15.871 &  0.013 &  0.015 &  0.021 &  0.184 &  0.124 &  0.135 &  0.010 &  A-Bl\\
   V3 &  198.214083 &   18.129278 & 0.63060 &  RRab & 16.004 & 15.734 & 15.679 &  0.016 &  0.015 &  0.021 &  0.378 &  0.300 &  0.312 &  0.011 &   A\\
   V4 &  198.182833 &   18.124000 & 0.38560 &   RRc & 16.102 & 15.902 & 15.868 &  0.014 &  0.016 &  0.021 &  0.182 &  0.120 &  0.147 &  0.012 &  A-Bl\\
   V5 &  198.162833 &   18.095194 & 0.63940 &  RRab & 15.945 & 15.687 & 15.647 &  0.015 &  0.015 &  0.023 &  0.481 &  0.352 &  0.328 &  0.010 &   A\\
   V6 &  198.266250 &   18.172194 & 0.66400 &  RRab & 15.932 & 15.649 & 15.616 &  0.017 &  0.017 &  0.028 &  0.446 &  0.314 &  0.295 &  0.007 &   A\\
   V7 &  198.253583 &   18.191667 & 0.54480 &  RRab & 16.119 & 15.852 & 15.844 &  0.020 &  0.018 &  0.026 &  0.468 &  0.284 &  0.327 &  0.011 &   A\\
   V8 &  198.251708 &   18.184750 & 0.61550 &  RRab & 16.030 & 15.777 & 15.736 &  0.016 &  0.018 &  0.025 &  0.482 &  0.327 &  0.324 &  0.008 &   A\\
   V9 &  198.250292 &   18.156972 & 0.60030 &  RRab & 16.041 & 15.780 & 15.730 &  0.019 &  0.021 &  0.029 &  0.447 &  0.303 &  0.325 &  0.008 &   A\\
  V10 &  198.190500 &   18.182083 & 0.60830 &  RRab & 16.013 & 15.764 & 15.702 &  0.016 &  0.018 &  0.023 &  0.489 &  0.342 &  0.304 &  0.029 &   A\\
  V11 &  198.189083 &   18.150556 & 0.62990 &  RRab & 15.975 & 15.714 & 15.677 &  0.013 &  0.014 &  0.019 &  0.449 &  0.317 &  0.293 &  0.012 &  A-Bl\\
  V12 &  198.349333 &   18.221278 & 0.61260 &  RRab & 15.970 & 15.718 & 15.701 &  0.021 &  0.020 &  0.025 &  0.498 &  0.318 &  0.331 &  0.013 &  A\\
  V13 &  198.367667 &   18.086972 & 0.62740 &  RRab & 15.971 & 15.681 & 15.648 &  0.026 &  0.021 &  0.045 &  0.453 &  0.332 &  0.322 &  0.056 &  A\\
  V14 &  198.335625 &   18.111861 & 0.54540 &  RRab & 16.075 & 15.826 & 15.812 &  0.016 &  0.016 &  0.029 &  0.514 &  0.338 &  0.350 &  0.040 &  A\\
  V15 &  198.301583 &   18.232028 & 0.30870 &   RRc & 16.303 & 16.130 & 16.098 &  0.014 &  0.015 &  0.023 &  0.147 &  0.092 &  0.083 &  0.009 &  A-Bl\\
  V16 &  198.192458 &   18.110917 & 0.30320 &   RRc & 16.317 & 16.149 & 16.118 &  0.017 &  0.018 &  0.026 &  0.176 &  0.125 &  0.112 &  0.016 &  A-Bl\\
  V17 &  198.168083 &   18.198361 & 0.38110 &   RRc & 16.158 & 15.957 & 15.919 &  0.017 &  0.015 &  0.022 &  0.195 &  0.137 &  0.137 &  0.042 &  A-Bl\\
  V18 &  198.202500 &   18.170333 & 0.33610 &   RRc & 16.271 & 16.076 & 16.040 &  0.018 &  0.018 &  0.025 &  0.242 &  0.134 &  0.132 &  0.021 &  A-Bl\\
  V19 &  198.279208 &   18.157333 & 0.39100 &   RRc & 16.130 & 15.907 & 15.871 &  0.021 &  0.020 &  0.028 &  0.173 &  0.128 &  0.128 &  0.010 &   A\\
  V20 &  198.287833 &   18.071194 & 0.38420 &   RRc & 16.120 & 15.903 & 15.888 &  0.019 &  0.018 &  0.029 &  0.177 &  0.105 &  0.095 &  0.029 &   A\\
  V21 &  198.358625 &   18.162028 & 0.33850 &   RRc & 16.239 & 16.032 & 15.986 &  0.021 &  0.021 &  0.033 &  0.160 &  0.107 &  0.094 &  0.037 &  A\\
  V23 &  198.259750 &   18.143333 & 0.36580 &   RRc & 16.148 & 15.935 & 15.912 &  0.020 &  0.020 &  0.030 &  0.205 &  0.133 &  0.108 &  0.007 &  A-Bl\\
  V24 &  198.197000 &   18.159000 & 0.76320 &  RRab & 15.802 & 15.524 & 15.469 &  0.014 &  0.015 &  0.019 &  0.243 &  0.268 &  0.246 &  0.013 &   A\\
  V25 &  198.268375 &   18.177028 & 0.70510 &  RRab & 15.926 & 15.633 & 15.582 &  0.019 &  0.017 &  0.025 &  0.333 &  0.280 &  0.285 &  0.006 &  A-Bl\\
  V26 &  198.149000 &   18.088917 & 0.39110 &   RRc & 16.096 & 15.875 & 15.846 &  0.016 &  0.017 &  0.025 &  0.162 &  0.151 &  0.102 &  0.013 &  A\\
  V27 &  198.172625 &   18.123278 & 0.67110 &  RRab & 15.928 & 15.665 & 15.611 &  0.013 &  0.014 &  0.019 &  0.403 &  0.310 &  0.306 &  0.008 &   A\\
  V28 &  198.175500 &   18.277139 & 0.63280 &  RRab & 15.966 & 15.705 & 15.676 &  0.018 &  0.015 &  0.025 &  0.471 &  0.368 &  0.307 &  0.042 &  A\\
  V29 &  198.267750 &   18.146389 & 0.82330 &  RRab & 15.771 & 15.458 & 15.421 &  0.019 &  0.018 &  0.028 &  0.192 &  0.185 &  0.186 &  0.006 &  A-Bl\\
  V30 &  198.250708 &   18.034444 & 0.53550 &  RRab & 16.145 & 15.903 & 15.889 &  0.021 &  0.025 &  0.026 &  0.303 &  0.214 &  0.187 &  0.027 &  A\\
  V31 &  198.248208 &   18.168028 & 0.70570 &  RRab & 15.922 & 15.647 & 15.588 &  0.020 &  0.017 &  0.023 &  0.508 &  0.338 &  0.334 &  0.009 &  A-Bl\\
  V32 &  198.198875 &   18.143306 & 0.39040 &   RRc & 16.085 & 15.865 & 15.826 &  0.015 &  0.016 &  0.022 &  0.171 &  0.112 &  0.117 &  0.010 &  A-Bl\\
  V33 &  198.182750 &   18.170306 & 0.62460 &  RRab & 16.007 & 15.744 & 15.695 &  0.014 &  0.014 &  0.020 &  0.445 &  0.316 &  0.330 &  0.030 &   A\\
  V34 &  198.190417 &   18.107278 & 0.28960 &   RRc & 16.357 & 16.206 & 16.166 &  0.014 &  0.017 &  0.024 &  0.093 &  0.064 &  0.097 &  0.019 &   A\\
  V35 &  198.260083 &   18.210528 & 0.37270 &   RRc & 16.133 & 15.911 & 15.866 &  0.014 &  0.017 &  0.023 &  0.200 &  0.136 &  0.146 &  0.011 &  A-Bl\\
  V36 &  198.263708 &   18.252889 & 0.37320 &   RRc & 16.169 & 15.959 & 15.914 &  0.015 &  0.015 &  0.023 &  0.164 &  0.109 &  0.099 &  0.015 &  A-Bl\\
  V37 &  198.217833 &   18.184833 & 0.71760 &  RRab & 15.907 & 15.623 & 15.562 &  0.018 &  0.017 &  0.024 &  0.412 &  0.304 &  0.304 &  0.018 &   A\\
  V38 &  198.238083 &   18.127944 & 0.70580 &  RRab & 15.880 & 15.622 & 15.564 &  0.015 &  0.014 &  0.021 &  0.321 &  0.301 &  0.277 &  0.007 &  A-Bl\\
  V40 &  198.232750 &   18.198528 & 0.31470 &   RRc & 16.341 & 16.152 & 16.117 &  0.017 &  0.019 &  0.026 &  0.128 &  0.096 &  0.096 &  0.015 &   A\\
  V41 &  198.236458 &   18.185694 & 0.61440 &  RRab & 16.050 & 15.796 & 15.741 &  0.019 &  0.017 &  0.025 &  0.405 &  0.306 &  0.249 &  0.015 &  A-Bl\\
  V42 &  198.210583 &   18.172250 & 0.71370 &  RRab & 15.876 & 15.584 & 15.551 &  0.018 &  0.016 &  0.021 &  0.380 &  0.298 &  0.246 &  0.011 &   A\\
  V43 &  198.221167 &   18.182083 & 0.71200 &  RRab & 15.892 & 15.614 & 15.568 &  0.018 &  0.018 &  0.024 &  0.347 &  0.251 &  0.260 &  0.018 &  A-Bl\\
  V44 &  198.215250 &   18.166833 & 0.37490 &   RRc & 16.183 & 15.989 & 15.928 &  0.022 &  0.022 &  0.028 &  0.192 &  0.139 &  0.098 &  0.014 &  A-Bl\\
  V45 &  198.229833 &   18.157611 & 0.65500 &  RRab & 15.945 & 15.675 & 15.628 &  0.027 &  0.026 &  0.029 &  0.442 &  0.326 &  0.249 &  0.010 &   A\\
  V46 &  198.227208 &   18.176944 & 0.70360 &  RRab & 15.910 & 15.609 & 15.584 &  0.018 &  0.020 &  0.022 &  0.311 &  0.279 &  0.222 &  0.016 &  A-Bl\\
  V47 &  198.210083 &   18.206861 & 0.33540 &   RRc & 16.191 & 16.004 & 15.997 &  0.018 &  0.017 &  0.025 &  0.138 &  0.102 &  0.113 &  0.026 &  A-Bl\\
  V51 &  198.239958 &   18.180472 & 0.35520 &   RRc & 16.155 & 15.933 & 15.909 &  0.021 &  0.025 &  0.032 &  0.157 &  0.075 &  0.086 &  0.006 &  A-Bl\\
  V52 &  198.233000 &   18.176972 & 0.37410 &   RRc & 16.229 & 16.002 & 15.968 &  0.021 &  0.024 &  0.028 &  0.172 &  0.091 &  0.107 &  0.013 &  A-Bl\\
  V53 &  198.232625 &   18.175444 & 0.38910 &   RRc & 16.044 & 15.610 & 15.519 &  0.025 &  0.028 &  0.030 &   ---  &   ---  &   ---  &  1.645 &  B-Bl\\
  V54 &  198.226292 &   18.175417 & 0.31510 &   RRc & 16.304 & 16.119 & 16.045 &  0.024 &  0.030 &  0.042 &  0.175 &  0.142 &  0.105 &  0.772 &  A-Bl\\
  V55 &  198.222750 &   18.176833 & 0.44330 &   RRc & 15.989 & 15.761 & 15.710 &  0.018 &  0.019 &  0.022 &  0.187 &  0.103 &  0.118 &  0.016 &  A-Bl\\
  V56 &  198.223708 &   18.157222 & 0.32890 &   RRc & 16.247 & 16.033 & 16.004 &  0.025 &  0.025 &  0.031 &  0.214 &  0.154 &  0.158 &  0.010 &   A\\
  V57 &  198.231500 &   18.166167 & 0.56830 &  RRab & 16.088 & 15.844 & 15.812 &  0.024 &  0.025 &  0.029 &  0.406 &  0.260 &  0.290 &  0.009 &  A-Bl\\
  V58 &  198.231667 &   18.158611 & 0.35500 &   RRc & 16.128 & 15.959 & 15.918 &  0.023 &  0.021 &  0.027 &  0.143 &  0.099 &  0.085 &  0.006 &  A-Bl\\
  V59 &  198.236083 &   18.155778 & 0.30390 &   RRc & 16.290 & 16.105 & 16.060 &  0.023 &  0.022 &  0.028 &  0.106 &  0.057 &  0.065 &  0.011 &  A-Bl\\
  V60 &  198.237417 &   18.160139 & 0.64480 &  RRab & 15.516 & 15.192 & 15.232 &  0.019 &  0.021 &  0.028 &   ---  &   ---  &   ---  &  0.016 &  B-Bl\\
  V61 &  198.229667 &   18.170139 & 0.37950 &   RRc & 15.844 & 15.461 & 15.357 &  0.038 &  0.055 &  0.059 &   ---  &   ---  &   ---  &  0.224 &  B-Bl\\
  V62 &  198.225000 &   18.174944 & 0.35990 &   RRc & 16.178 & 16.000 & 15.900 &  0.084 &  0.108 &  0.110 &   ---  &   ---  &   ---  &  0.027 &  B-Bl\\
  V63 &  198.234542 &   18.166861 & 0.31050 &   RRc & 16.257 & 16.069 & 16.033 &  0.021 &  0.022 &  0.029 &  0.122 &  0.076 &  0.106 &  0.407 &   A\\
  V64 &  198.218833 &   18.170139 & 0.31970 &   RRc & 14.745 & 14.270 & 14.198 &  0.015 &  0.016 &  0.017 &   ---  &   ---  &   ---  &  0.013 &   B\\
  V71 &  198.226583 &   18.165000 & 0.30450 &   RRc & 16.343 & 16.100 & 16.100 &  0.051 &  0.069 &  0.087 &   ---  &   ---  &   ---  &  0.019 &   B\\
  V72 &  198.233083 &   18.164528 & 0.34070 &   RRc & 16.128 & 16.115 & 16.069 &  0.025 &  0.030 &  0.033 &  0.100 &  0.094 &  0.096 &  0.498 &   A\\
  V91 &  198.223417 &   18.170444 & 0.30240 &   RRc & 16.406 & 16.205 & 16.126 &  0.024 &  0.022 &  0.028 &  0.106 &  0.083 &  0.102 &  0.014 &   A\\
  V92 &  198.228792 &   18.180639 & 0.27720 &   RRc & 16.436 & 16.251 & 16.222 &  0.021 &  0.022 &  0.028 &  0.097 &  0.109 &  0.139 &  0.015 &   A\\
\enddata
\tablecomments{Star ID, coordinates (epoch J2000), periods, and subtypes are taken from \citet{clement2001}. $\Delta$ is the separation between the coordinates of RR Lyrae
from \citet{clement2001} and our astrometry. Quality flags (QF) - `A' represents well sampled light curves while `B' refers to blended targets; `Bl' indicates known Blazhko variation.}
\end{deluxetable*}

\begin{figure}
\epsscale{1.2}
\plotone{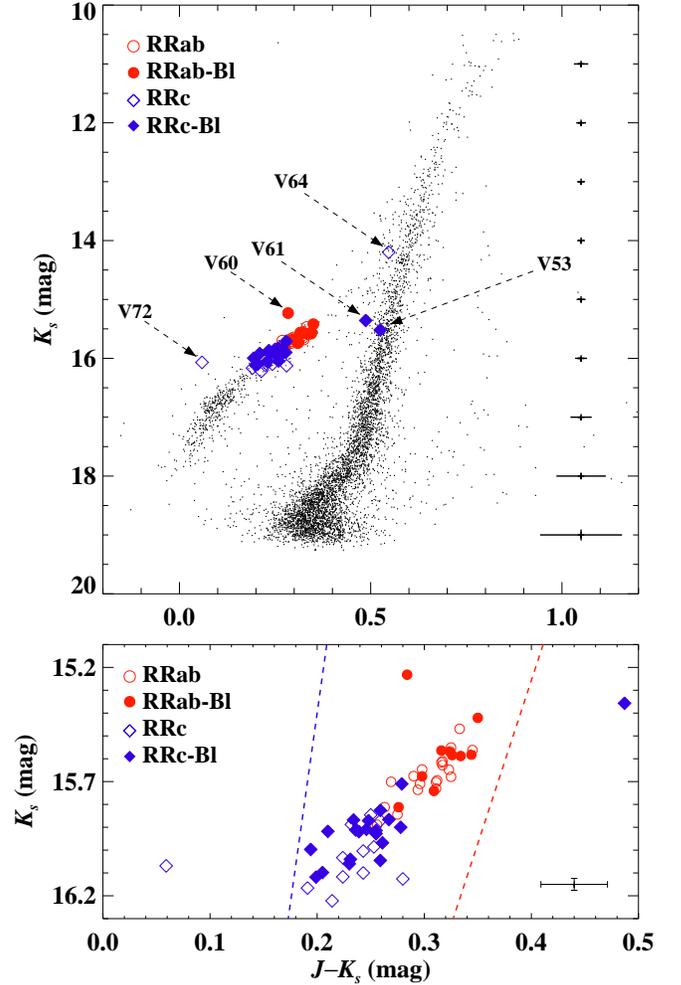}
\caption{{\it Top:} Color--magnitude diagram for stars in M53 with three-band photometry. RR Lyrae variables are also shown. Some RR Lyrae that appear to be outliers among horizontal branch variables are also marked (see text for details). Representative $\pm2\sigma$ error bars in both magnitudes and colors are shown. {\it Bottom:} Zoom-in on the horizontal branch RR Lyrae. The solid blue and dashed red lines display the predicted first overtone blue edge and the fundamental red edge from \citet{marconi2015}, respectively. Non-Blazhko/Blazhko RR Lyrae are shown as open/filled symbols. A representative median error bar is also shown.} 
\label{fig:cmd_all}
\end{figure}

\begin{figure}
\epsscale{1.2}
\plotone{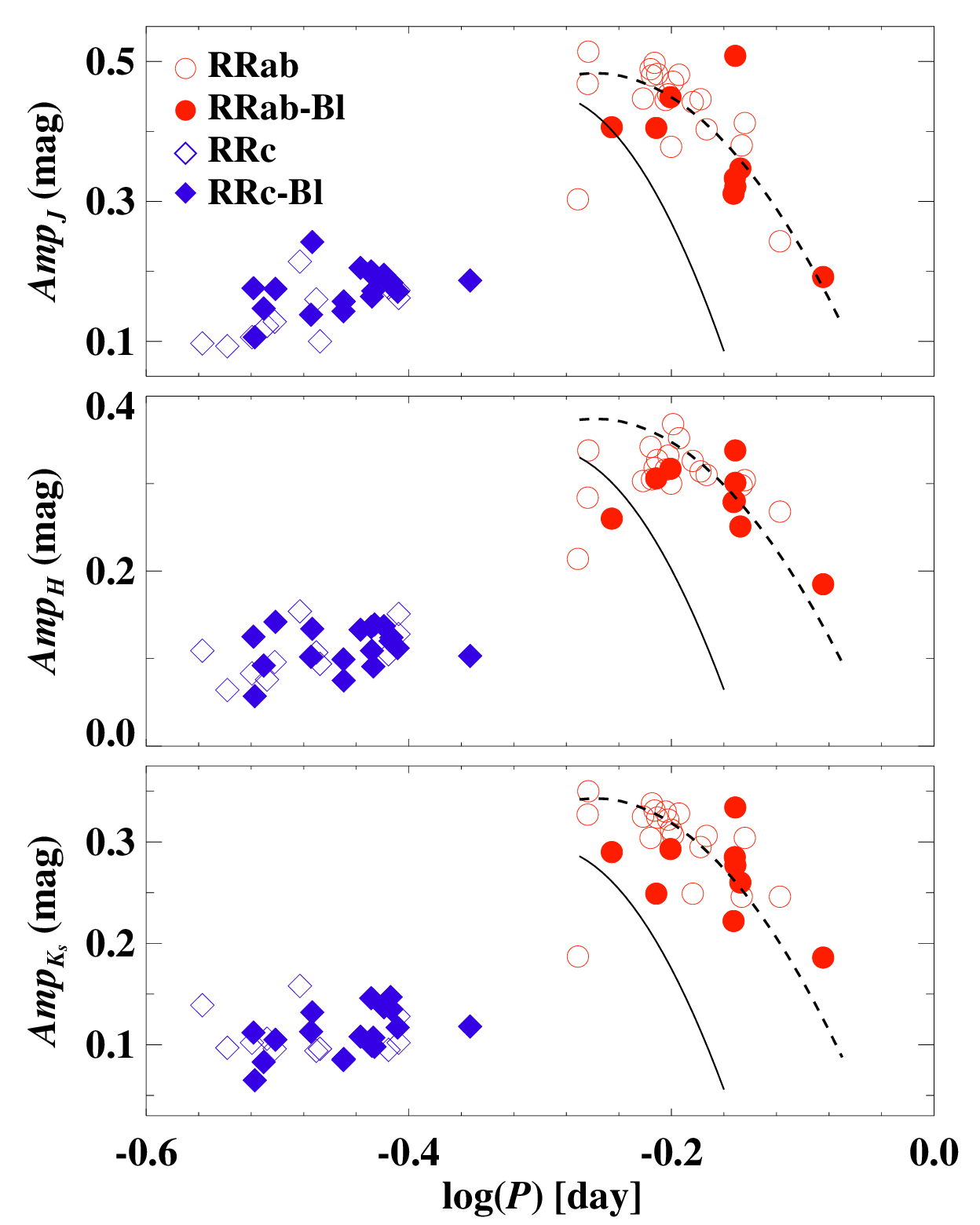}
\caption{Bailey diagrams for RR Lyrae stars in M53 in the $J$ (top), $H$ (middle), and $K_s$ (bottom) bands. The dashed lines display the arbitrarily scaled $I$-band loci for OoII RRab in M53 from \citet{ferro2011} as discussed in the text. The overplotted solid lines represent the approximate $JHK_s$ loci of OoI RRab in M3 \citep[][]{bhardwaj2020a}.}
\label{fig:bailey_jhk}
\end{figure}

The top panel of Fig.~\ref{fig:cmd_all} shows the proper motion cleaned color--magnitude diagram for M53 sources. RR Lyrae located along the horizontal branch are also shown, while those found to be outliers with respect to their expected location on the horizontal branch, either due to different color or magnitude are labeled. All these RR Lyrae also have peculiar positions in the optical color--magnitude diagrams due to photometric blending \citep[see Figure 4 in][]{ferro2011}, except V72. While V72 is significantly bluer, the apparent $K_s$-band magnitude for this RRc star is consistent with overtone variables on the horizontal branch. Two RR Lyrae (V60 and V64) are also outliers in the proper motions along the declination axis. Similarly, V54 is a proper motion outlier but located perfectly within the RR Lyrae horizontal branch in the color--magnitude diagram. However, these three variables (V54, V60, and V64) are located in the unresolved central $1'$ region of the cluster and are likely blended with nearby brighter stars.

The bottom panel of Fig.~\ref{fig:cmd_all} shows the horizontal branch RR Lyrae and the predicted boundaries of the instability strip. The analytical relations for the edges of the instability strip
in NIR bands are independent of the adopted metal-abundances \citep{marconi2015}. A distance modulus of 16.4 mag \citep{ferro2011} was used to offset the theoretically predicted red and blue edges of the instability strip into the observational color--magnitude diagram. There is a good agreement between the theoretical and observed blue edge whereas the empirical red edge appears bluer than the predicted one. This discrepancy might be ascribed to the dependence of the fundamental red edge on the efficiency of super-adiabatic convection. If a slightly higher convective efficiency were assumed, a bluer theoretical red edge would be obtained \citep[see for e.g.,][]{dicriscienzo2004}. 

\citet{ferro2012} noted a clear separation between RRab and RRc in optical color--magnitude diagrams. At NIR wavelengths, there is little overlap between RRab and RRc in the $(J-K_s)$, $K_s$ color--magnitude diagram and all RR Lyrae with good photometric quality flags are located within the theoretically predicted instability strip. \citet{ferro2012} suggested that the horizontal branch evolution in M53 probably occurs toward the red edge leading to a clear separation between fundamental and first-overtone mode pulsators. Furthermore, RRc displaying Blazhko variations are redder and brighter than stable RRc variables suggesting that their modulations are connected to changes in the pulsation mode. 

Fig.~\ref{fig:bailey_jhk} displays the period--amplitude or Bailey \citep{bailey1902} diagrams for M53 RR Lyrae in the $JHK_s$-bands for the first time. The locus of OoII type M53 RRab from \citet{ferro2011} in the $I$-band was arbitrarily scaled by 65\%, 50\%, and 45\% to fit the $J$, $H$, and $K_s$-band Bailey diagrams, respectively. The loci of the OoI cluster M3 from \citet{bhardwaj2020a} are also shown for comparison. The majority of RRab follow the expected trend for OoII type clusters. The smallest period RRab (V30) seems to have a smaller amplitude which is most likely due to larger scatter in the light curves around the minima and maxima. \citet{ferro2012} found that Blazhko RRc stars have relatively larger $V$-band amplitudes albeit with a larger scatter presumably due to amplitude modulations. At NIR wavelengths, no obvious trend is seen in the amplitudes for Blazhko and non-Blazhko stars. This suggests that the amplitude modulations are not significant at NIR wavelengths as noted by \citet{jurcsik2018} who found that Blazhko modulation is primarily driven by the change in the temperature variation.

\section{Period--Luminosity relations} 
\label{sec:RR Lyrae_plr}

\begin{figure*}
\epsscale{1.2}
\plotone{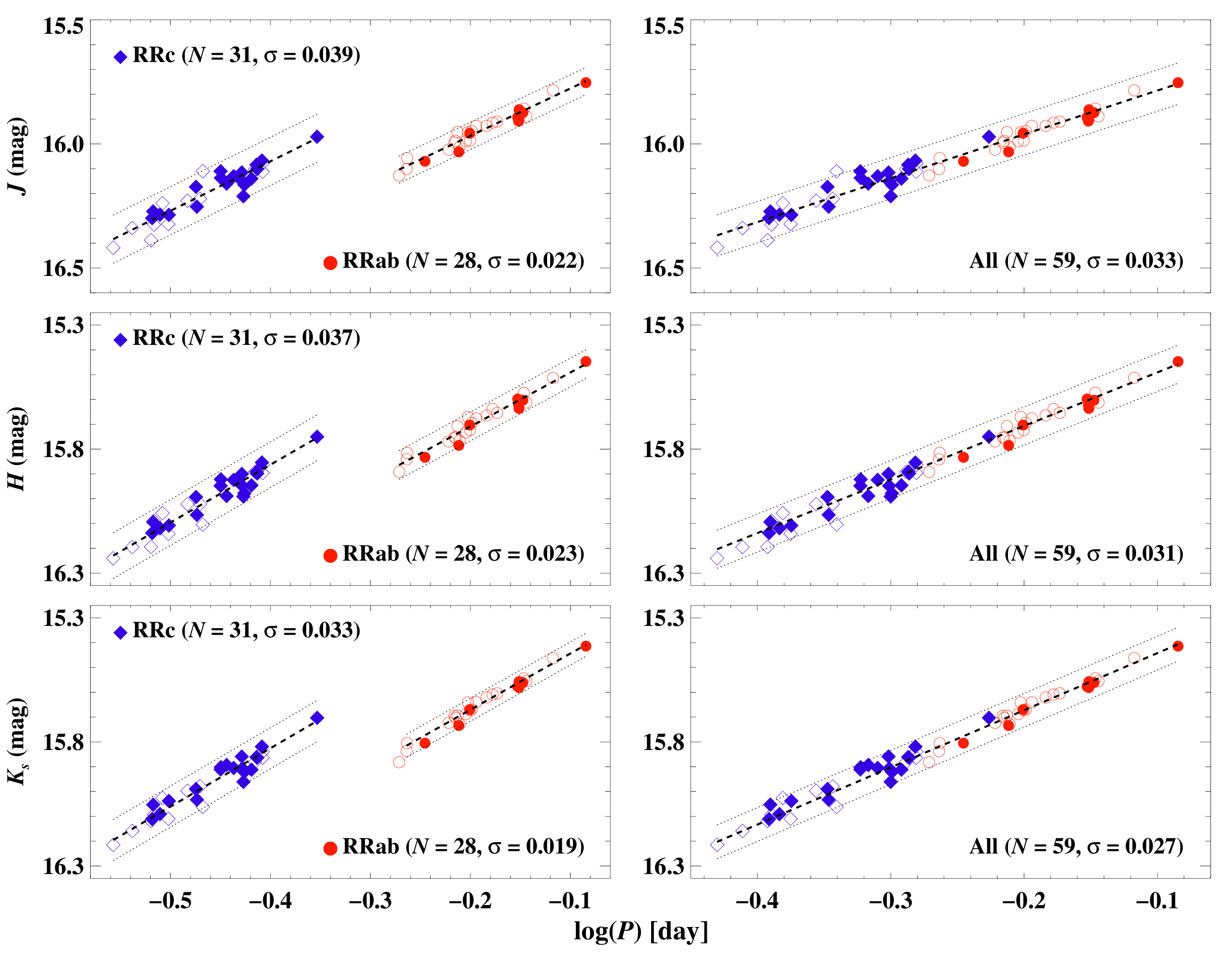}
\caption{NIR PLRs for M53 RRab and RRc (left) and all RR Lyrae (right) in $J$ (top), $H$ (middle), and $K_s$ (bottom). 
The dashed lines represent best-fitting linear regression over the period range under consideration while the dotted lines display $\pm 3\sigma$ offsets from the best-fitting PLRs.} 
\label{fig:nir_plr}
\end{figure*}

Mean magnitudes for RR Lyrae in the $JHK_s$-bands were used to derive PLRs of the following form:
\begin{equation}
{m_\lambda} = a_\lambda + b_\lambda\log(P),
\label{eq:plr}
\end{equation}

\noindent where $a_\lambda$ and $b_\lambda$ are the slope and zero-point of the PLR in a given filter. The interstellar reddening in M53 is small, $E(B-V) = 0.02$~mag \citep{harris2010}. Adopting 
total-to-selective absorption ratios from \citet{bhardwaj2017a} and the reddening law of \citet{card1989} assuming $R_V=3.1$, extinction corrections of 18, 11, and 7 mmag were estimated in the 
$J$, $H$, and $K_s$-bands, respectively. Three different samples of RR Lyrae (RRab, RRc, and RRab+RRc) were considered to derive their PLRs. To combine the sample of RR Lyrae, periods of RRc stars were 
fundamentalized using the equation: $\log(P_{\textrm{FU}})=\log(P_{\textrm{FO}})+0.127$ \citep{petersen1991, coppola2015}, where `FU' and `FO' represent fundamental and first-overtone modes, respectively. 

\begin{deluxetable}{rrrrrr}
\tablecaption{NIR PLRs of RR Lyrae in the M53 cluster. \label{tbl:plr_nir}}
\tabletypesize{\footnotesize}
\tablewidth{0pt}
\tablehead{\colhead{Band} & \colhead{Type} & \colhead{$b_\lambda$} & \colhead{$a_\lambda$} & \colhead{$\sigma$}& \colhead{$N$}\\}
\startdata
     $J$ &  RRab &    15.585$\pm$0.021      &     $-1.910\pm$0.110      &      0.022 &   28\\
     $J$ &   RRc &    15.272$\pm$0.046      &     $-1.995\pm$0.100      &      0.039 &   31\\
     $J$ &   All &    15.608$\pm$0.011      &     $-1.768\pm$0.042      &      0.033 &   59\\
     $H$ &  RRab &    15.273$\pm$0.021      &     $-2.177\pm$0.111      &      0.023 &   28\\
     $H$ &   RRc &    14.931$\pm$0.047      &     $-2.329\pm$0.102      &      0.037 &   31\\
     $H$ &   All &    15.275$\pm$0.011      &     $-2.158\pm$0.045      &      0.031 &   59\\
   $K_s$ &  RRab &    15.217$\pm$0.032      &     $-2.267\pm$0.173      &      0.019 &   28\\
   $K_s$ &   RRc &    14.885$\pm$0.065      &     $-2.352\pm$0.141      &      0.033 &   31\\
   $K_s$ &   All &    15.212$\pm$0.016      &     $-2.303\pm$0.063      &      0.027 &   59\\
\enddata
\tablecomments{The zero-point ($b$), slope ($a$), dispersion ($\sigma$) and 
the number of stars ($N$) in the final PLR fits are tabulated.}
\end{deluxetable}

In the color--magnitude diagram shown in Fig.~\ref{fig:cmd_all}, three RRc (V53, V61, and V64) and one RRab (V60) have significantly brighter magnitudes than the horizontal branch RR Lyrae stars.
Therefore, these RR Lyrae were excluded from further analysis and the resulting sample of 59 stars is used to derive PLRs. Although another RR Lyrae (V72) is significantly bluer than the horizontal branch variables, its apparent magnitudes are consistent with those of RRc stars. Fig.~\ref{fig:nir_plr} displays $JHK_s$-band PLRs for different samples of RR Lyrae in M53. The best-fitting PLRs are listed in Table~\ref{tbl:plr_nir}. The slopes of the PLRs show the typical trend of being shallower for RRab compared with RRc stars while the slopes of the global sample including both RRab and RRc are the shallowest \citep[e.g.,][]{braga2018, beaton2018, bhardwaj2020a}. The NIR PLRs listed in Table~\ref{tbl:plr_nir} are statistically robust against a different threshold for outlier removal than the adopted $3\sigma$ clipping and statistically consistent for different samples of Blazhko and non-Blazhko RR Lyrae variables. For RRab stars, the PLRs exhibit unprecedentedly small scatter ($\sigma \sim 0.02$~mag) that is comparable to the median photometric uncertainties in the mean magnitudes. This is expected firstly because no significant metallicity spread \citep[$\sigma_{\textrm{[Fe/H]}}=0.07$~dex,][]{boberg2016} is noted from high-resolution spectra of bright red giant branch stars in M53. Secondly, the $JHK_s$ light curves of RR Lyrae in M53 are well-sampled because their periods are not close to 0.5 days thus leading to accurate and precise determinations of their mean-magnitudes. Furthermore, no trend is seen in the residuals of the $JHK_s$-band PLRs. 

\begin{figure}
\epsscale{1.2}
\plotone{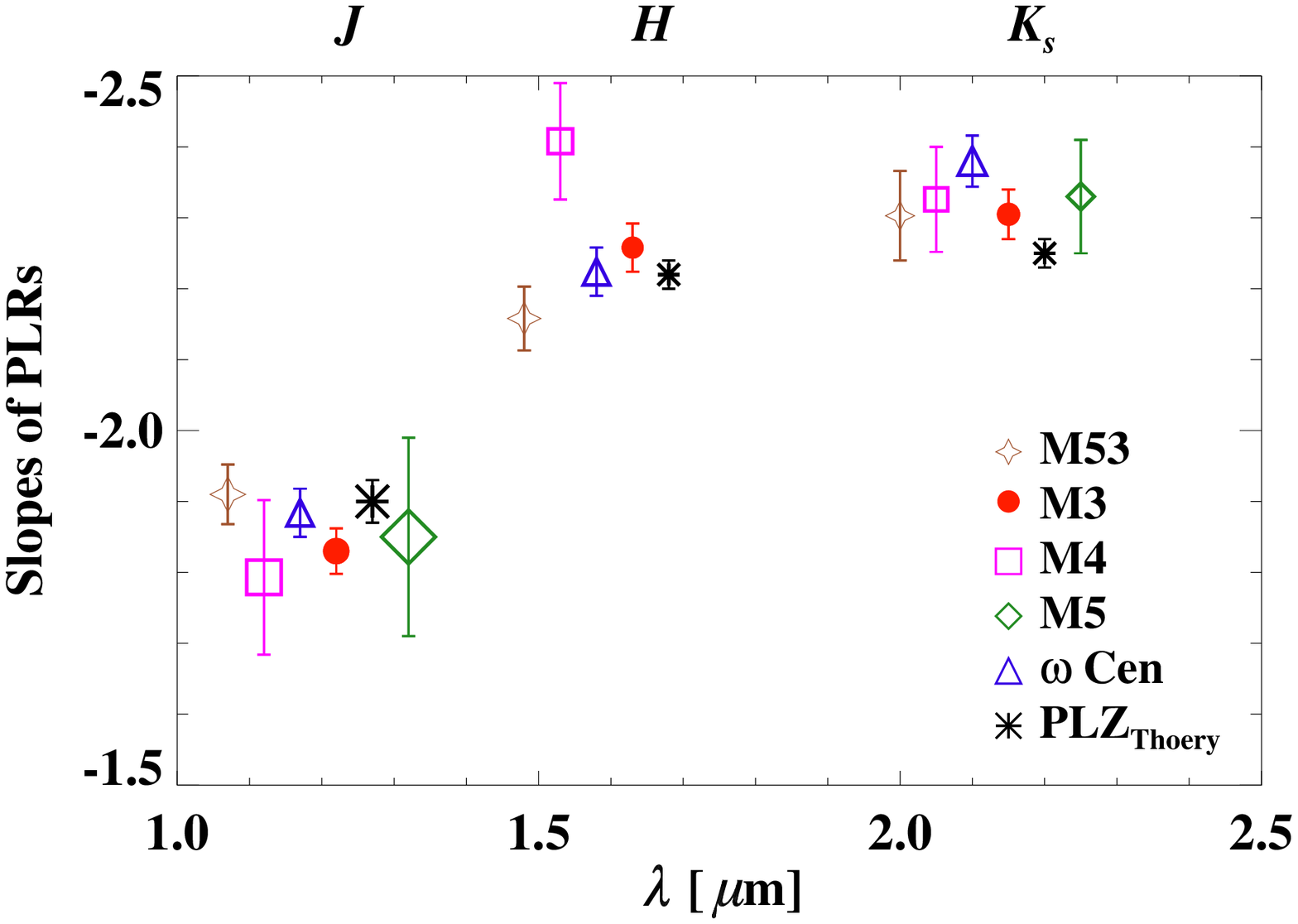}
\caption{The slopes of $JHK_s$-band PLRs of RR Lyrae variables in different GCs. Each data point corresponding to a PLR in a given filter is slightly shifted along the {\it x}-axis for visual clarity. A larger symbol size represents a larger dispersion in the underlying PLRs in a given GC. The empirical slopes were taken from:  M5 \citep{coppola2011}, M4 \citep{braga2015}, $\omega$ Cen \citep{braga2018}, M3 \citep{bhardwaj2020a}, and the theoretical slopes of PLZ relation were adopted from \citet{marconi2015}. }
\label{fig:slp_plr}
\end{figure}

Our empirical PLRs for the combined sample of RR Lyrae in M53 were found to be in agreement with empirical relations in other globular clusters \citep[M4, $\omega$ Cen, M3,][]{braga2015, braga2018, bhardwaj2020}, and with predicted relations based on non-linear pulsation models \citep{marconi2015} as shown in Fig.~\ref{fig:slp_plr}. While the slopes of our RRc PLRs are quantitatively shallower than the theoretical predictions ($J:-2.46,~H:-2.70,~K_s:-2.72$), the numerical values are still consistent within the quoted uncertainties. Comparing with an OoI cluster M3 \citep{bhardwaj2020a}, the PLRs for M53 RR Lyrae are marginally shallower but also consistent within the errors in all three NIR filters. 

\subsection{Distance to the M53 cluster}
\label{sec:sec_3.1}

M53 is one of the most distant Milky Way GCs and precise independent distances to this cluster are lacking in the literature. Table~\ref{tbl:mu_m53} lists a few recent estimates of distance moduli to M53 based on different methods. \citet{dekany2009} estimated a true distance modulus to M53 using the optical ($VI$)-band PLC relation for RR Lyrae variables. The authors used a zero-point calibration based on Baade-Wesselink (BW) analysis of 21 stars \citep{kovacs2003} which suffers from systematics, for example, in the so-called projection factor that is adopted to convert the observed radial velocity into a pulsation velocity \citep{gallenne2017}. Distance estimates based on the visual magnitude--metallicity relation for RR Lyrae also suffer from systematics due to evolutionary effects \citep{bono2003} and photometric metallicity estimates typically have large uncertainties. \citet{ferro2011} adopted a non-standard approach to determine absolute $V$-band magnitudes for RR Lyrae from light-curve Fourier decomposition. Since extinction corrections are significant at optical wavelengths, these distance moduli are also sensitive to the adopted reddening and the extinction law even though the reddening toward M53 is small.  

The catalog of \citet{harris2010} lists an apparent visual distance modulus of 16.32~mag to the M53 cluster. \citet{wagner2016} used this distance modulus as input prior and performed Bayesian analyses to fit stellar isochrones. The authors estimated several cluster parameters including the distance modulus which varies between 16.35 and 16.48~mag depending on whether they fitted single or double stellar population isochrones, and adopted helium and $\alpha$-element abundances \citep[see Tables 3 and 6 of][]{wagner2016}.  Recently, \citet{hernitschek2019} determined distances to several GCs including M53 by modeling the spatial distribution of RRab stars within and near overdensities in their sample GCs. From Table~\ref{tbl:mu_m53}, the distance estimates to M53 range from 18 to 19.8 kpc but the uncertainties in these measurements are only the standard error on the mean values. The systematic uncertainties in different methods are not taken into account in these measurements.  

\begin{deluxetable}{cclc}
\tablecaption{Distance to the M53 cluster. \label{tbl:mu_m53}}
\tabletypesize{\footnotesize}
\tablewidth{0pt}
\tablehead{\colhead{$\mu$} & \colhead{[Fe/H]} & \colhead{Method} & \colhead{Ref.}\\
	mag		&	 		&		 	&	\\}
\startdata
16.31$\pm$0.04	&	---		&	RR Lyrae optical PLC relation		&D09	\\
16.35$\pm$0.11	&	$-1.58\pm$0.03	&	$M_V$--[Fe/H]$^a$ relation		&D09	\\
16.48$\pm$0.14	&	$-2.12\pm$0.05	&	$M_V$--[Fe/H]$^b$ relation		&D09	\\
16.36$\pm$0.05	&	$-1.92\pm$0.06	&	$M_V$ RRab 				&A11	\\	
16.28$\pm$0.07  &	$-1.92\pm$0.06	&	$M_V$ RRc 				&A11	\\
16.39$\pm$0.01	&	$-2.10\pm$0.05	&	Stellar isochrone fitting		&W16$^c$	\\	
16.47$\pm$0.01	&	$-2.10\pm$0.05	&	Stellar isochrone fitting		&W16$^d$	\\	
16.31$\pm$0.02	&	---		&	Density model fitting			&H19	\\
16.43$\pm$0.02  &       $-2.06\pm$0.05	&	RR Lyrae PLZ$_J$ relation 		&TW	\\
16.39$\pm$0.02  &       $-2.06\pm$0.05	&	RR Lyrae PLZ$_H$ relation		&TW	\\
16.40$\pm$0.02  &       $-2.06\pm$0.05	&	RR Lyrae PLZ$_{K_s}$ relation		&TW	\\
16.36$\pm$0.07	&	$-2.06\pm$0.05	&	PLZ$_{K_s}$ and Gaia EDR3 + $\pi_\mathrm{corr}^{Fix}$		&TW	\\
16.32$\pm$0.06	&	$-2.06\pm$0.05	&	PLZ$_{K_s}$ and Gaia EDR3 + $\pi_\mathrm{corr}^{L20}$		&TW	\\  
\hline
\multicolumn{3}{l}{$\mu_\textrm{M53} = 16.403~\pm~0.024$~(stat.)~$\pm~0.033$~(syst.)~mag}	&TW\\
\hline
\multicolumn{3}{l}{$D_\textrm{M53} = 19.081~\pm~0.211$~(stat.)~$\pm~0.290$~(syst.)~kpc}	&TW\\
\enddata
\tablecomments{D09 \citep{dekany2009} determined photometric metallicities for RR Lyrae ([Fe/H]$^a$) and red giants ([Fe/H]$^b$), respectively. A11 \citep{ferro2011} determined $M_V$ using an empirical relation based on Fourier parameters. The distance modulus from W16 \citep{wagner2016} is based on single (W16$^c$) and double (W16$^d$) population isochrone fitting to color--magnitude diagrams. H16 \citep{hernitschek2019} estimated their distance modulus by modeling the RR Lyrae overdensity and spatial distributions in GCs. TW - This work. For {\it Gaia} based estimates, two values correspond
to the parallax correction ($\pi_\mathrm{corr}^{Fix}=-7\pm3~\mu$as) based on predicted parallaxes and those from \citet[$\pi_\mathrm{corr}^{L20}$,][]{lindegren2020a}, respectively. $\mu_\textrm{M53}$ - adopted distance modulus to M53.}
\end{deluxetable}

NIR PLRs for RR Lyrae are excellent distance indicators and therefore can provide a very precise independent distance to M53. With our new robust mean $JHK_s$ magnitudes, the precision of estimated distances will only be limited by the uncertainties in the absolute calibration of RR Lyrae PLRs. In most recent studies, theoretical PLZ relations of RR Lyrae have been employed for distance measurement \citep{marconi2015, neeley2017, braga2018, bhardwaj2020a}. Empirical calibrations based on {\it Gaia} DR2 and {\it Hubble Space Telescope} parallaxes of RR Lyrae are known to exhibit larger systematic uncertainties \citep[see][for more details]{neeley2017, muraveva2018, bhardwaj2020a}. Therefore, we estimated a distance to M53 using two independent calibrations based on the theoretical PLZ relations and the latest {\it Gaia} EDR3.

First, we employed the theoretical PLZ relation for RR Lyrae in the $JHK_s$-bands from \citet{marconi2015} and adopted a mean iron-abundance of [Fe/H]$=-2.06$~dex for M53. This [Fe/H] value is taken from \citet{carretta2009} since the theoretical PLZ relation uses metallicities on their scale. The coefficients of the theoretical relations were used to anchor the zero-points for the empirical $JHK_s$-band PLRs listed in Table~\ref{tbl:plr_nir}. The results of distance moduli estimates using $JHK_s$-band PLRs for the combined sample of RR Lyrae are tabulated in Table~\ref{tbl:mu_m53}. For the sample of RRab or RRc, the distance moduli estimates are consistent with the quoted values within their $1\sigma$ uncertainties. However, the uncertainties on distance measurements based on RRc stars are larger due to the relatively larger scatter and errors in the slopes and zero-points of their PLRs. The systematic uncertainties were quantified by adding errors to the zero-points, uncertainties due to variations in the mean-metallicity, and errors in the slopes of the calibrator and M53 RR Lyrae PLRs. A distance modulus to the M53 cluster of $\mu = 16.403~\pm~0.024$~(stat.)~$\pm~0.033$~(syst.)~mag was obtained by taking a weighted mean of all measurements based on different samples of RR Lyrae. Our distance modulus to M53 is in good agreement with the most distances moduli estimates listed in Table~\ref{tbl:mu_m53}.

Secondly, we used parallaxes of Galactic RR Lyrae from recent {\it Gaia} EDR3 \citep{lindegren2020} to empirically calibrate the zero-point of NIR PLZ relations. The Galactic RR Lyrae sample of 403 stars was adopted from the compilation of \citet{dambis2013} which includes periods, RR Lyrae subtype, spectroscopic metallicites, $K_s$-band magnitudes, and the extinction in the $V$-band. \citet{dambis2013} applied random-phase corrections to $K_s$ measurements obtaining mean-magnitudes for most stars; 32 stars in their sample have only single-epoch $K_s$-band magnitudes. \citet{muraveva2018} used the same sample for an investigation of RR Lyrae as distance indicators using {\it Gaia} DR2 parallaxes, and also updated periods and subtypes of a few variables. The authors also found a significant parallax zero-point offset of $-57~\mu$as for this RR Lyrae sample in the {\it Gaia} DR2, which was nearly two times larger than the quasar based value of $-29~\mu$as \citep{lindegren2018}. In {\it Gaia} EDR3, \citet{lindegren2020} found a parallax bias of $-17~\mu$as using quasars. 

\begin{figure}
\epsscale{1.2}
\plotone{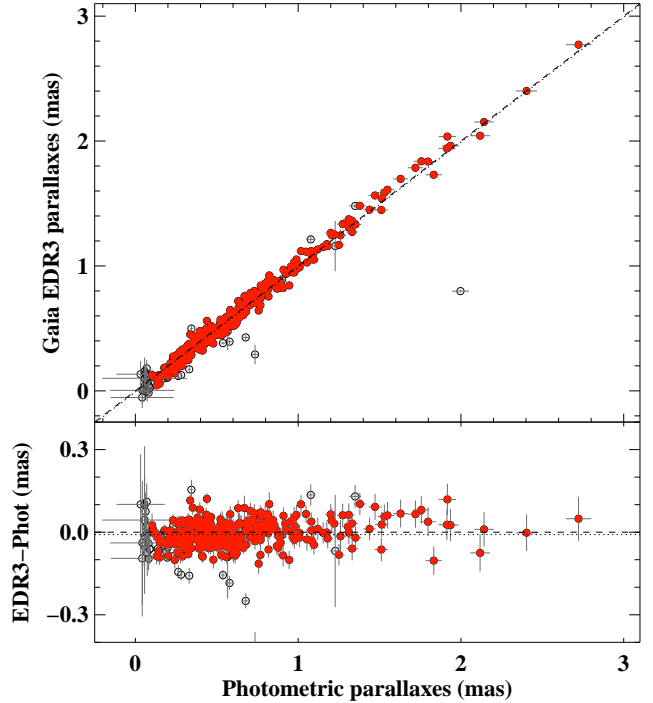}
\caption{{\it Top:} Comparison of {\it Gaia} EDR3 parallaxes for a sample of 403 stars with predicted photometric parallaxes. The dashed line represents $y=x$ and the dotted line shows the offset between the two sets of parallax measurements. {\it Bottom:} Difference between {\it Gaia} EDR3 and photometric parallaxes. RR Lyrae (350 of 403 stars) with filled red symbols are those with (1) positive parallaxes; (2) re-normalized unit weight error (RUWE) $<2$, (3) $K_s \leq 16$~ mag, (4) random-phase corrected $K_s$ measurements, and (5) residuals within $3\sigma$ of the best-fitting line.}
\label{fig:plx_off}
\end{figure}

\begin{figure}
\epsscale{1.2}
\plotone{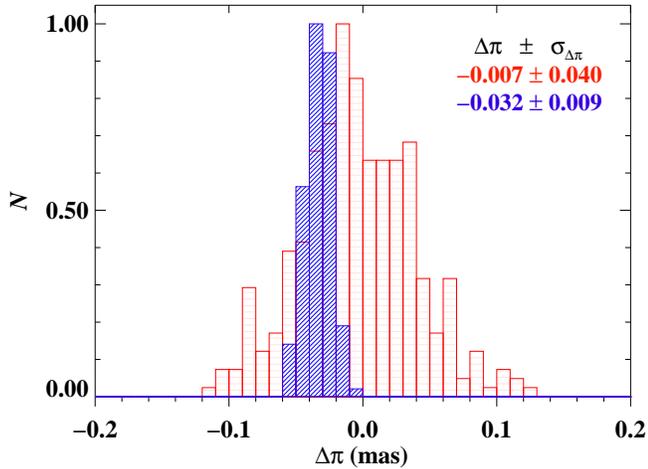}
\caption{Normalized histogram of parallax zero-point offsets estimated using the theoretical PLZ$_{K_s}$ relation. The parallax bias values determined using the recipe suggested by \citet{lindegren2020a} are also shown using slanted blue lines. The peak values and the standard deviations of the two sets of zero-point offsets are also shown.}
\label{fig:plx_hist}
\end{figure}

We used theoretical PLZ relations in the $K_s$-band to determine absolute magnitudes and then predict parallaxes for the full sample of RR Lyrae from \citet{dambis2013} or \citet{muraveva2018}. For this purpose, extinction corrections were applied to the $K_s$-band apparent magnitudes and the spectroscopic metallicites, which are on the \citet{zinn1984} metallicity scale, were transformed to the \citet{carretta2009} scale to be consistent with theoretical models. {\it Gaia} EDR3 data were obtained for all 403 RR Lyrae. Three stars have negative parallaxes. Fig.~\ref{fig:plx_off} shows a comparison of {\it Gaia} EDR3 parallaxes with predicted photometric parallaxes. The two sets of parallaxes exhibit a strong correlation with a minimal residual offset consistent with zero. 

Few selection criteria (see caption of Fig.~\ref{fig:plx_off}) were applied to select 350 out of 403 stars for which the histogram of parallax offsets between {\it Gaia} EDR3 and photometric parallaxes is shown in Fig.~\ref{fig:plx_hist}. We found a small residual parallax bias, $-7\pm3~\mu$as ($\sigma=40~\mu$as), statistically consistent with zero within $3\sigma$ and given the large scatter, in the sense that the {\it Gaia} EDR3 parallaxes are systematically smaller. However, the difference between the predicted and EDR3 parallaxes exhibits a dependence on the size of parallaxes. It becomes significant ($-17\pm3~\mu$as) for small parallaxes ($<0.5~$mas) but is consistent with zero for larger parallaxes  ($\geq 0.5~$mas). \citet{marconi2020} also used theoretical optical period-wesenheit relations to predict parallaxes for these RR Lyrae in {\it Gaia} DR2 and found a negligible small zero-point offset. For a relative comparison, the median difference in parallaxes between {\it Gaia} EDR3 and DR2 for our RR Lyrae sample amounts to $21\pm3~\mu$as, which also hints at a smaller zero-point offset in the improved EDR3 parallaxes. 

We also estimated the parallax zero-point offset using the recipe provided by \citet{lindegren2020a} for correcting the EDR3 parallaxes. The normalized histogram of parallax offsets estimated using their method is also shown as slanted blue lines in Fig.~\ref{fig:plx_hist}. It peaks at a statistically significant value of $-32\pm4~\mu$as exhibiting a very small ($9~\mu$as) scatter. However, these parallax corrections depend in a non-trivial way on the magnitude, colour, and ecliptic latitude of the source, and are difficult to quantify for bright variable sources like RR Lyrae. The median parallax offsets for the two sets of data plotted in the Fig.~\ref{fig:plx_hist} are more consistent for faint stars with small parallaxes. The difference is mostly for bright sources, and it could be due to the lack of bright LMC calibrator stars with effective wavenumber ($\sim 1.5-1.65~\mu\rm{m^{-1}}$) similar to our RR Lyrae stars \citep[see Figure 9 of][]{lindegren2020a}. 

We compared {\it Gaia} EDR3 parallaxes that included \citet{lindegren2020a} parallax-correction recipe with the parallaxes predicted by the theoretical PLZ relation, similar as was done in Fig.~\ref{fig:plx_off} for the no parallax-correction case. The slope (1.030$\pm$0.003 dex) of the best-fitting line is similar to the no parallax-correction case but the intercept becomes consistent with zero. This suggests that the parallax corrections are more significant for the stars with the smaller parallax, i.e. predominantly fainter RR Lyrae. However, the median difference between the two sets of parallaxes becomes positive, $22\pm2~\mu$as ($\sigma=39~\mu$as), and significantly larger than the no parallax-correction case ($-7\pm3~\mu$as), which points towards an over-correction in parallaxes for our sample of RR Lyrae in {\it Gaia} EDR3 data. Using \citet{lindegren2020a} formulae, \citet{riess2021} also found an indication of over-correction of the parallax offset for Cepheids with a median difference of $15~\mu$as in the residuals.  

Finally, we used a smaller subsample of 251 Galactic RR Lyrae stars by applying several selection criteria: (1) $\sigma_{\pi}/{\pi} \leq 0.05$; (2) $\pi > 0$; (3) re-normalized unit weight error (RUWE) $<2$, (4) $K_s \leq 16$~ mag, (5) random-phase corrected $K_s$ measurements. Although these cuts can bias our sample, we also note that the compilation of \citet{dambis2013} is already limited to stars with spectroscopic metallicities and is possibly biased toward brighter sources ($>90\%$ stars are RRab). \citet{bailer2020} suggested that for stars with positive parallaxes and $\sigma_{\pi}/{\pi} \leq 0.1$, the inverse parallax is a reasonably good distance estimate. Therefore, we used the parallax distances by applying a $7~\mu$as parallax offset calibrating the absolute zero-point in $K_s$-band for the M53 PLRs and estimate a distance modulus, $\mu_{\rm M53} = 16.366~\pm~0.042$~(stat.)~$\pm~0.067$~(syst.)~mag, consistent with the distance estimated using theoretical PLZ relations. When correcting the parallaxes using the recipe by \citet{lindegren2020a}, we find a distance modulus, $\mu_{\rm M53} = 16.320~\pm~0.039$~(stat.)~$\pm~0.058$~(syst.)~mag, to M53. We further discuss the possible systematics resulting from limiting the sample to precise parallaxes in the next section by providing a comparison with the results based on \citet{bailer2020} distance catalog for {\it Gaia} EDR3 sources.

\section{Constraints on the metallicity dependence}
\label{sec:plzr}

New NIR time-series data in GCs of different Oosterhoff types and mean-metallicities can be used to constrain the metallicity effect on the RR Lyrae PLRs. Theoretical models of RRab predict a metallicity coefficient of $\sim 0.18$ mag dex$^{-1}$ in the $K_s$-band \citep{marconi2015}. If the independent distances are known to two clusters with $\Delta\textrm{[Fe/H]}= 1.0$ dex then the magnitudes for an RRab with a given period are expected to differ, on average, by 0.18 mag. However, this relative difference in luminosity between clusters exhibiting a significant metallicity difference is difficult to quantify due to the internal dispersion of RR Lyrae PLRs, a lack of precise independent distances, and a possible spread in their mean-metallicities. 

We compiled NIR mean-magnitudes for RR Lyrae variables from time-series data in the M5 \citep{coppola2011}, M4 \citep{stetson2014}, $\omega$ Cen \citep{braga2018}, and M3 \citep{bhardwaj2020a} GCs. Homogeneous $JHK_s$-band magnitudes are available for all these clusters, except for M5 which has only $JK_s$ NIR photometry \citep{coppola2011}. Combining PLRs in these GCs with the new M53 data we construct a sample of five GCs with high quality PLRs for RR Lyrae which allows us to empirically investigate the metallicity dependence of PLRs in the NIR. Since the $J$-band light curves of RR Lyrae in M5 exhibit uneven phase coverage and no templates were used to determine their mean-magnitudes \citep[see,][]{coppola2011}, only $K_s$-band PLRs are considered. Note that we only consider clusters for which multi-epoch RR Lyrae photometry is available. 

\begin{deluxetable*}{lcrrrrr}
\tablecaption{$K_s$-band PLZ relation and the estimated distances to the GCs. \label{tbl:mu_gcs}}
\tabletypesize{\footnotesize}
\tablewidth{0pt}
\tablehead{\colhead{~~~~~~~~~~~~~~~~~~~~~~~~~~~~~~~~~~~} & \colhead{} & \colhead{M4} & \colhead{M5} 	&\colhead{M3} 	& \colhead{$\omega$ Cen}	& \colhead{M53}	}
\startdata
              $N$		&     	 	&                   43 &                  76 &                 220 &                  74$^a$ &                59 \\
         $E(B-V)$		&   	mag	&                0.370 &               0.030 &               0.013 &               0.110 &             0.020  \\
           {[Fe/H]}		&   	dex	&       $-1.18\pm0.02$ &      $-1.33\pm0.02$ &      $-1.50\pm0.05$ &      $-1.64\pm0.09$ &    $-2.06\pm0.09$  \\
   $\mu_{\rm RRL}$		&   	mag	&       11.27$\pm$0.01 &      14.44$\pm$0.02 &      15.04$\pm$0.04 &      13.67$\pm$0.04 &    16.40$\pm$0.03  \\
             Ref.		&      		&                  B15 &                 C11 &                 B20 &                 B18 &                TW  \\
   $\mu_{\rm Other}$		&   	mag	&       11.30$\pm$0.05 &      14.34$\pm$0.09 &      15.07$\pm$0.01 &      13.51$\pm$0.12 &    16.39$\pm$0.01  \\
             Ref.		&      		&                  K13 &                 G19 &                 T19 &                 K07 &               W16  \\
\hline
\multicolumn{7}{c}{No zero-point calibration \citep[{[Fe/H]} for $\omega$ Cen RR Lyrae from][]{magurno2019}}\\
\hline
            $a_i$		&   	mag	&    10.352$\pm$0.019  &    13.357$\pm$0.018 &    14.076$\pm$0.018 &    12.715$\pm$0.019 &    15.338$\pm$0.018  \\
        \multicolumn{7}{l}{$b=    -2.332\pm0.006     $}\\
        \multicolumn{7}{l}{$c=     0.071\pm0.012     $}\\
\hline
\multicolumn{7}{c}{No zero-point calibration \citep[{[Fe/H]} for $\omega$ Cen RR Lyrae from][]{sollima2006}}\\
\hline
            $a_i$		&   	mag	&    10.440$\pm$0.018  &    13.464$\pm$0.020 &    14.198$\pm$0.018 &    12.828$\pm$0.019 &    15.501$\pm$0.017  \\
        \multicolumn{7}{l}{$b=    -2.309\pm0.007     $}\\
        \multicolumn{7}{l}{$c=     0.145\pm0.012     $}\\
\hline
\multicolumn{7}{c}{Theoretical PLZ calibration from \citet{marconi2015}}\\
\hline
  \multicolumn{7}{l}{$a_{cal}=    -0.842\pm0.005     $}\\
          $\mu_i$		&   	mag	&    11.323$\pm$0.013  &    14.342$\pm$0.014 &    15.079$\pm$0.013 &    13.718$\pm$0.014 &    16.399$\pm$0.013  \\
            $a_i$		&   	mag	&    10.481$\pm$0.012  &    13.500$\pm$0.013 &    14.237$\pm$0.012 &    12.876$\pm$0.013 &    15.557$\pm$0.012  \\
        \multicolumn{7}{l}{$b=    -2.312\pm0.004     $}\\
        \multicolumn{7}{l}{$c=     0.170\pm0.009     $}\\
\hline
\multicolumn{7}{c}{Calibration using {\it Gaia} EDR3 distances from \citet{bailer2020} which include parallax offsets from \citet{lindegren2020a}}\\
\hline
  \multicolumn{7}{l}{$a_{cal}=    -0.766\pm0.017     $}\\
          $\mu_i$		&   	mag	&    11.241$\pm$0.025  &    14.257$\pm$0.026 &    14.995$\pm$0.025 &    13.633$\pm$0.026 &    16.311$\pm$0.025  \\
            $a_i$		&   	mag	&    10.475$\pm$0.019  &    13.491$\pm$0.020 &    14.229$\pm$0.018 &    12.867$\pm$0.019 &    15.545$\pm$0.018  \\
        \multicolumn{7}{l}{$b=    -2.311\pm0.010     $}\\
        \multicolumn{7}{l}{$c=     0.164\pm0.015     $}\\
\hline
\multicolumn{7}{c}{Calibration using {\it Gaia} EDR3 parallaxes from \citet{lindegren2020} including a fixed parallax offset ($-7\pm3~\mu$as)}\\
\hline
  \multicolumn{7}{l}{$M_{cal}=    -0.836\pm0.016     $}\\
          $\mu_i$		&   	mag	&    11.292$\pm$0.026  &    14.310$\pm$0.027 &    15.043$\pm$0.025 &    13.678$\pm$0.026 &    16.351$\pm$0.024  \\
            $a_i$		&   	mag	&    10.456$\pm$0.020  &    13.474$\pm$0.022 &    14.207$\pm$0.019 &    12.842$\pm$0.020 &    15.515$\pm$0.018  \\
        \multicolumn{7}{l}{$b=    -2.318\pm0.007     $}\\
        \multicolumn{7}{l}{$c=     0.151\pm0.013     $}\\
\hline
\multicolumn{7}{c}{Theoretical PLZ + {\it Gaia} EDR3 parallaxes based calibration}\\
\hline
  \multicolumn{7}{l}{$a_{cal}=    -0.848\pm0.007     $}\\
          $\mu_i$		&   	mag	&    11.312$\pm$0.017  &    14.340$\pm$0.018 &    15.071$\pm$0.017 &    13.699$\pm$0.018 &    16.388$\pm$0.017  \\
            $a_i$		&   	mag	&    10.464$\pm$0.015  &    13.492$\pm$0.017 &    14.223$\pm$0.016 &    12.851$\pm$0.017 &    15.540$\pm$0.015  \\
        \multicolumn{7}{l}{$b=    -2.320\pm0.006     $}\\
        \multicolumn{7}{l}{$c=     0.166\pm0.011     $}\\
\enddata
\tablecomments{$N$ is the number of RR Lyrae in each GC considered in this analysis. [Fe/H] values are adopted from \citet{carretta2009} and $E(B-V)$ values are taken from \citet{harris2010}. $^a$For 
$\omega$ Cen, individual spectroscopic metallicities for 125 RR Lyrae from \citet{magurno2019} were considered for the first case and those of 74 stars from \citet{sollima2006} were considered for other cases. In the no calibration case, zero-points ($a_i$), slope ($b$), and metallicity ($c$) coefficients were obtained after solving Equation~(\ref{eq:eqnk}) while in other cases the coefficients are determined by solving Equation~(\ref{eq:eqnk1}). $\mu_i$ are the distance moduli and $a_{cal}$ is the calibrated zero-point of the PLZ$_{K_s}$ relation. Literature distance moduli (DM) based on RR Lyrae PLRs ($\mu_{\textrm{~RRL}}$) and other independent methods ($\mu_{\textrm{~Other}}$) are taken from the following references: K07 - \citet{kaluzny2007}, C11 - \citet{coppola2011}, K13 - \citet{kaluzny2013},  B15 - \citet{braga2015}, W16 - \citet{wagner2016}, B18 - \citet{braga2018}, G19 - \citet{gontcharov2019},   T19 - \citet{tailo2019}, B20 - \citet{bhardwaj2020a}, TW - this work.}
\end{deluxetable*}

To constrain the metallicity dependence, mean [Fe/H] values for the GCs were taken from \citet{carretta2009} adopting a homogeneous set of GC metallicity measurements on the same scale as the theoretical models. RR Lyrae in $\omega$ Cen  exhibit a significant spread in metallicity and therefore, we consider two samples of RR Lyrae with spectroscopic metallicities from \citet{sollima2006a} and \citet{magurno2019}. The former provide spectroscopic metallicities for 74 RR Lyrae while the latter increased the sample size to 125 stars in $\omega$ Cen. Since the number of RRab in M4 and M53 is significantly smaller compared with other GCs, the combined sample of RR Lyrae (RRab+RRc) is used in this analysis. 

The apparent $K_s$-band magnitudes of RR Lyrae were corrected for extinction using the color-excess values listed in Table~\ref{tbl:mu_gcs} combined with the adopted extinction law of \citet{card1989} assuming the standard ratio of absolute to selective extinction, $R_V=3.1$. However, \citet{hendricks2012} suggested a larger value of $R_V=3.6$ for M4 which is used to estimate the $K_s$-band extinction for RR Lyrae in this GC. Although there is evidence of differential reddening in M4 \citep[$\sigma_{E(B-V)}\geq 0.2$~mag,][]{hendricks2012} and also in $\omega$ Cen \citep[$\sigma_{E(B-V)}\sim0.03$~mag,][]{calamida2005}, only mean reddening is considered for all RR Lyrae in each cluster as adopted in \citet{braga2015} and \citet{braga2018} for M4 and $\omega$ Cen, respectively. 

\subsection{The $K_s$-band Period--Luminosity--Metallicity relation}

We considered the following cases in deriving the empirical PLZ$_{K_s}$ relation for RR Lyrae in our sample GCs:

\subsubsection{No zero-point calibration} First, the $K_s$-band PLZ relation was derived using a single slope and metallicity coefficient by fitting an equation of the following form:

\begin{eqnarray}
\label{eq:eqnk}
m_{{i,j}} & = & a_{i} + b\log P_{i,j} + c\textrm{[Fe/H]}_{i}, 
\end{eqnarray}

\noindent where $m_{i,j}$ are the extinction corrected $K_s$-band magnitudes for the $j^{\rm th}$ RR Lyrae in the $i^{\rm th}$ GC. The coefficients $a_{i}$ represent different zero-points of the PLZ relation while $b$ and $c$ are the common slope and metallicity coefficients. We solved Equation~(\ref{eq:eqnk}) by constructing a matrix-formalism of dependent and independent parameters and using $\chi^2$ minimization as discussed in detail by \citet[][their equation 5]{bhardwaj2016a}. For a robust analysis, an iterative $3\sigma$ clipping was used to remove the single largest outlier in each iteration until convergence. 

We obtained a universal slope and different zero-points for each GC which are listed in Table~\ref{tbl:mu_gcs}. A smaller but statistically significant metallicity coefficient, $c = 0.071\pm0.012$, was obtained using more recent spectroscopic metallicities for 125 $\omega$ Cen RR Lyrae from \citet{magurno2019}. However, this metallicity dependence is at least two times smaller than the theoretical predictions of \citet{marconi2015}. Using the spectroscopic metallicities for 74 $\omega$ Cen RR Lyrae from \citet{sollima2006a}, a significant metallicity coefficient, $c = 0.145\pm0.012$, was obtained which is consistent with the results of \citet{navarrete2017} and \citet{braga2018} only based on $\omega$ Cen data. The variation in the metallicity coefficients could be attributed to the difference in spectroscopic metallicities between the two samples which is significant ($\Delta$[Fe/H]$=-$0.13 dex) for stars in common, and is likely due to the adopted techniques used to determine metallicities \citep{magurno2019}. The peak of the metallicity distribution from \citet{sollima2006} is on average, $\sim 0.2$ dex more metal-rich than that from \citet{magurno2019}. Furthermore, the \citet{sollima2006} sample shows a metal-rich tail as a distinct secondary peak in the metallicity distribution which is not present in the \citet{magurno2019} sample. We only considered the \citet{sollima2006} sample for the further analysis.

\subsubsection{Theoretical calibration} Following studies on RR Lyrae distance scales \citep[see][for a review]{beaton2018, bhardwaj2020}, we adopted the theoretical PLZ$_{K_s}$ relation for RR Lyrae to calibrate the zero-point of the empirically derived PLZ relation in the GCs. The theoretical models were taken from \citet{marconi2015} for a wide range of metallicities ($0.1 < {\rm [Fe/H]} < -2.4$~dex). We fitted an equation of the following form, simultaneously solving for the absolute zero-point and the distances to the GCs:  

\begin{eqnarray}
\label{eq:eqnk1}
m_{{i,j}} & = & a_{\rm cal} + \mu_i + b\log P_{i,j} + c\textrm{[Fe/H]}_{i}, 
\end{eqnarray}

\noindent where $\mu_i = a_i - a_{\rm cal}$ and the other coefficients are the same as in Equation~(\ref{eq:eqnk}). The calibrated zero-point ($a_{\rm cal}$) and distances to each GC along with other coefficients are listed in Table~\ref{tbl:mu_gcs}. Since we solved using theoretical models together with PLRs in the GCs, the metallicity coefficient is more consistent with the predicted value of $0.18$~mag dex$^{-1}$. Furthermore, the distance moduli are also in agreement with the literature values based on RR Lyrae and other independent methods (see Table~\ref{tbl:mu_gcs}).

\subsubsection{{\it Gaia} EDR3 calibration} 

Finally, we also used {\it Gaia} EDR3 parallaxes for the sample of \citet{dambis2013} to calibrate the PLZ$_{K_s}$ relation for RR Lyrae in our GCs. To avoid any biases due to restrictions on the precision of parallaxes, we initially obtained distances to all RR Lyrae in our sample from the latest catalog of \citet{bailer2020}. The authors adopted a probabilistic approach to estimating stellar distances that uses a prior constructed from a three-dimensional model of the Milky Way. This approach allows reasonable distance estimates even for stars with negative parallaxes and large parallax uncertainties for which a simple inversion of parallax is not recommended. Note that the catalog of \citet{bailer2020} provides distances after applying the parallax zero-point correction derived by \citet{lindegren2020a}. We solved Equation~(\ref{eq:eqnk1}) adopting the calibration based on these {\it Gaia} EDR3 distances and the results are listed in Table~\ref{tbl:mu_gcs}. The slope and metallicity coefficients are consistent with the results based on the theoretical calibrations but the calibrated zero-point is significantly smaller thus leading to relatively smaller distance moduli estimates to the GCs.

To obtain a robust zero-point calibration, we restricted the sample to 251 stars with criteria discussed in Section~\ref{sec:sec_3.1}. We emphasize that while these cuts may lead to possible biases if a PLZ relation is determined only based on {\it Gaia} EDR3 data. However, our analysis includes strong constraints on the slope of the PLZ relation thanks to precise photometric data in the GCs. Therefore, only precise parallaxes were selected to obtain a robust zero-point calibration. Furthermore, we only applied a fixed parallax zero-point offset of $-7\pm3~\mu$as as determined previously. We used a simple inversion of parallaxes for this subsample of 251 stars and the estimated distances in Equation~(\ref{eq:eqnk1}) for calibrating the PLZ$_{K_s}$ relation. The results based on the calibration using {\it Gaia} EDR3 parallaxes in Table~\ref{tbl:mu_gcs} clearly show that the coefficients and distance moduli are now in good agreement with those based on the theoretical calibrations. This further validates that the parallax zero-point offset for our sample of Galactic RR Lyrae estimated using the theoretical PLZ relations is small and negligible within the uncertainties. However, it is also possible that the theoretical calibrations lead to systematically larger distance moduli estimates within $1-2\sigma$ of the quoted uncertainties which negates the parallax zero-point offset when comparing {\it Gaia} and photometric parallaxes.

\begin{figure*}
\epsscale{1.2}
\plotone{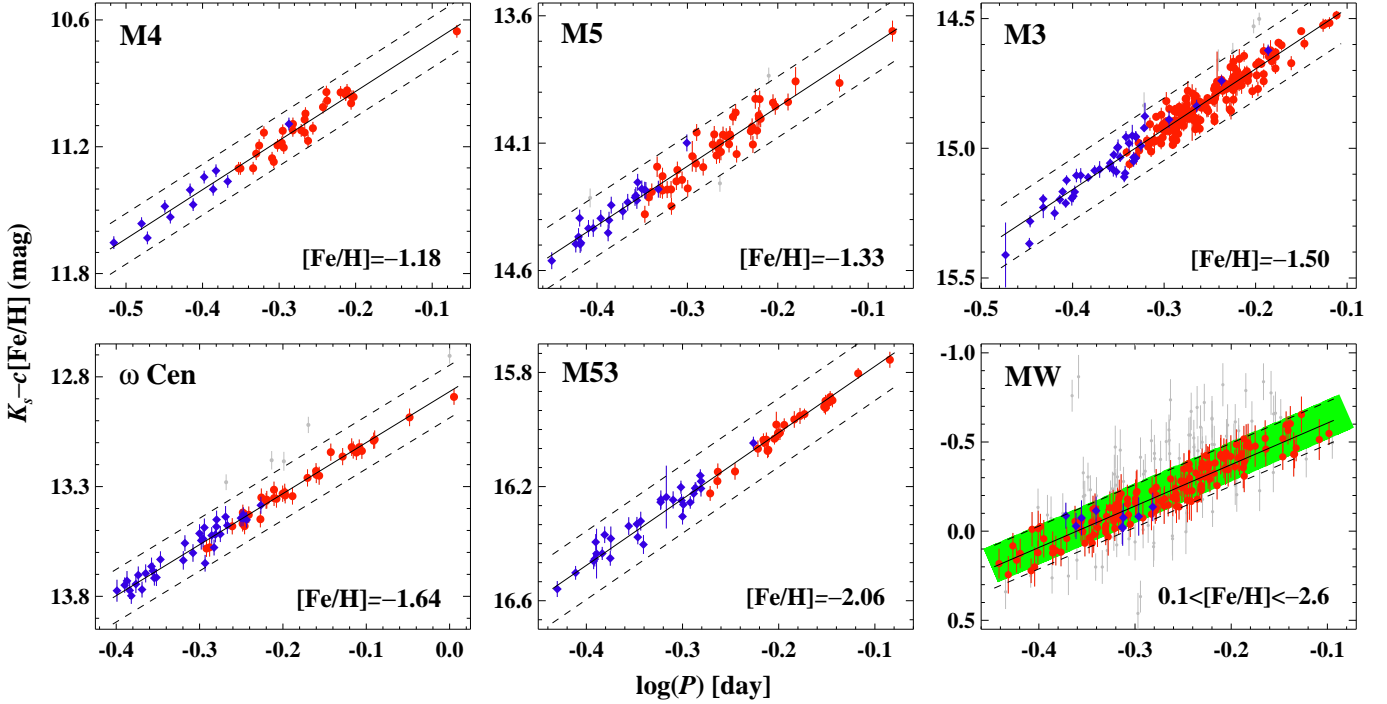}
\caption{PLZ$_{K_s}$ relation for RR Lyrae in different GCs. The calibrator PLZ$_{K_s}$ relation for Galactic RR Lyrae is also shown in the bottom right panel. The green shaded region shows the calibrator theoretical relation of \citet{marconi2015} with $\pm3\sigma$ scatter. Solid lines display the best-fitting common slope of the PLZ$_{K_s}$ relation to all GCs while dashed lines represent $\pm 3\sigma$ scatter in the underlying relation. Mean-metallicities for each cluster are also indicated in each panel. Red circles and blue squares represent RRab and (fundamentalized) RRc stars, respectively. Grey symbols represent $3\sigma$ outliers.}
\label{fig:plrs_k}
\end{figure*}

\begin{figure}
\epsscale{1.2}
\plotone{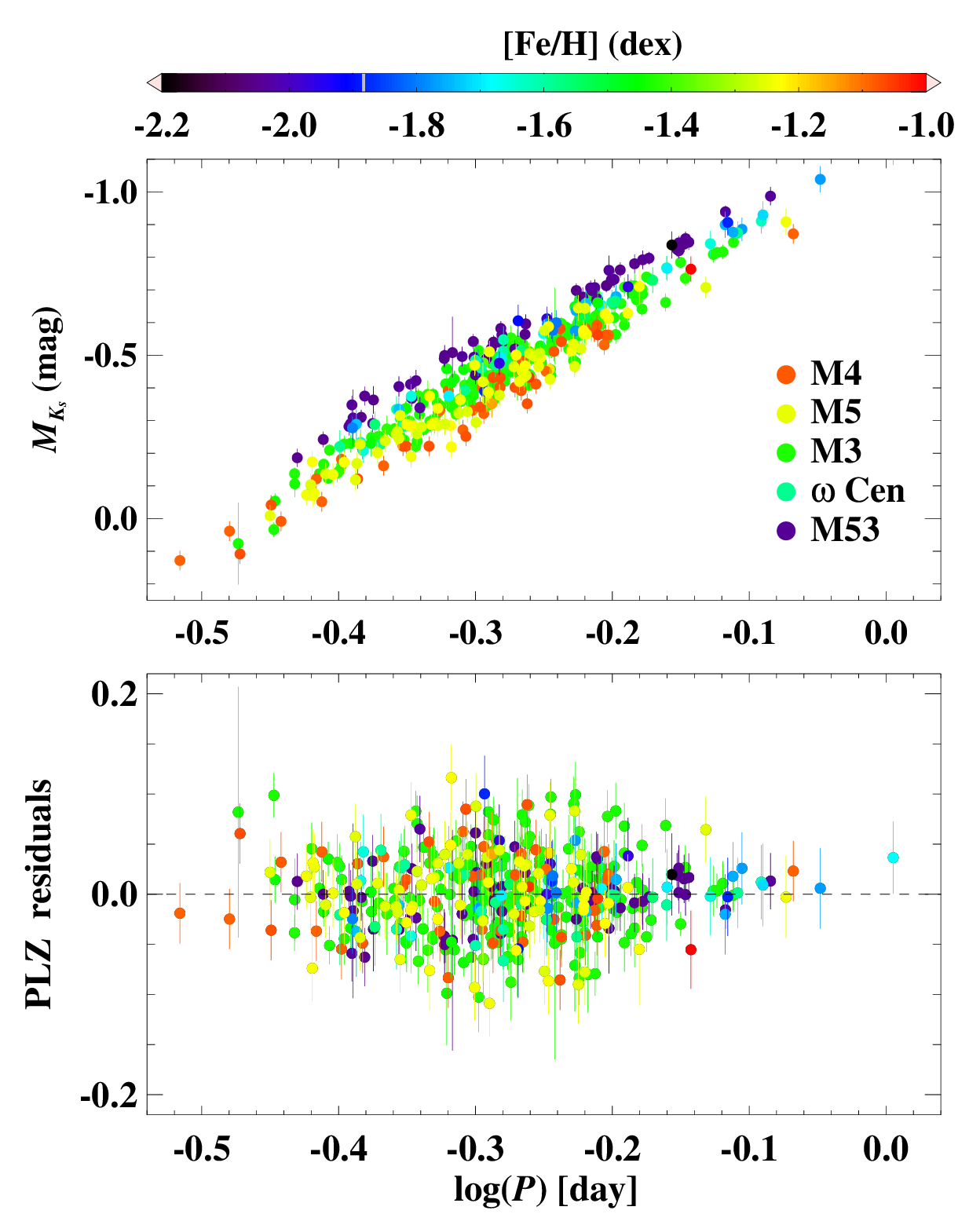}
\caption{{\it Top:} Combined $K_s$-band PLR for RR Lyrae in different GCs based on their absolute magnitudes determined using the distances listed in Table~\ref{tbl:mu_gcs}. For $\omega$ Cen, the color code in legend corresponds to mean-metallicity but its RR Lyrae stars exhibit scatter in [Fe/H]. {\it Bottom:} Residuals of the PLZ$_{K_s}$ relation for RR Lyrae from Equation~(\ref{eq:eqnk}). Note the metallicity trend in the PLR in the top panel which disappears in the residuals of the PLZ$_{K_s}$ relation.}
\label{fig:plzk}
\end{figure}

Finally, we adopted the calibration derived using Equation~(\ref{eq:eqnk1}) including both theoretical models and the subsample of {\it Gaia} EDR3 parallaxes. In this case, we did not apply any zero-point offset assuming a $-7\pm3~\mu$as bias is not statistically significant, given the scatter, as discussed in the previous section. Therefore, an absolute zero-point calibration was obtained purely based on the best-fitting relation to the absolute magnitudes from both the theoretical models and the {\it Gaia} EDR3 data. Fig.~\ref{fig:plrs_k} displays the common PLZ$_{K_s}$ relation for RR Lyrae in different GCs and the results are listed in Table~\ref{tbl:mu_gcs}. Our distance estimates to these GCs, except for M5, are consistent with the RR Lyrae PLR based estimates in the literature. The distance moduli to these GCs are also in agreement with independent measurements either based on eclipsing binaries \citep{kaluzny2007, kaluzny2013} or isochrone-fitting to color--magnitude diagrams \citep{wagner2016, gontcharov2019, tailo2019}. For M5, the distance modulus is similar to the value estimated by \citet{sollima2006} based on the PLZ$_{K_s}$ relation for RR Lyrae variables. The difference in the M5 distance modulus with respect to \citet{coppola2011} could be attributed to their adopted theoretical calibration, the different mean-metallicity value ([Fe/H]$\sim-1.26$~dex), and the uncertainties in the transformations from 2MASS to the Johnson--Cousins--Glass photometric system. Fig.~\ref{fig:plzk} shows the calibrated PLR for RR Lyrae based on their absolute magnitudes determined using the distance moduli estimates listed in Table~\ref{tbl:mu_gcs}. The PLRs in the different GCs exhibit the expected trend as a function of the metal-abundance with RR Lyrae in metal-rich clusters being fainter. The residuals of the PLZ$_{K_s}$ relation in Equation~(\ref{eq:eqnk1}), shown in the bottom panel of Fig.~\ref{fig:plzk}, do not exhibit any metallicity dependence. 

\section{Discussion and Conclusions} \label{sec:discuss}

We have presented new PLRs for RR Lyrae in the M53 GC for the first time at NIR wavelengths. RR Lyrae multi-epoch photometry was obtained using the WIRCam instrument on the CFHT covering a $21'\times 21'$ region around the cluster center. New NIR data were complemented with accurate pulsation periods from the updated catalog of \citet{clement2001} and precise and accurate mean-magnitudes were determined by fitting templates to the well-sampled $JHK_s$-band light curves. The resulting PLRs for our RR Lyrae exhibit an unprecedentedly small scatter for RRab with a dispersion of $\sim0.02$~mag in all three filters, which is comparable to the precision of our photometry for the faintest variable. Multi-epoch data were also used to investigate color--magnitude and Bailey diagrams for M53 RR Lyrae at NIR wavelengths. No significant overlap is seen in the color--magnitude plane between RRab and RRc stars suggesting that horizontal branch evolution occurs toward the red edge of the instability strip \citep{ferro2012}. RR Lyrae in the Bailey diagrams follow the expected locus of RRab in a typical OoII clusters. 

New NIR photometry for RR Lyrae in the metal-poor M53 GC provided an opportunity to constrain the metallicity dependence of their PLRs by means of a quantitative comparison with relatively metal-rich GCs. While single-epoch data for RR Lyrae are available for several GCs, for example from 2MASS, homogeneous $JHK_s$-band time-series photometry is limited to only a few GCs. Time-series data are crucial to determine precise mean magnitudes and the PLRs for RR Lyrae with a dispersion of $\sim 0.04$~mag, thus allowing a rigorous comparison of GCs with different mean-metallicities. Combining precise $K_s$-band PLRs in GCs covering a metallicity range of $\Delta\textrm{[Fe/H]}=0.9$~dex, we determined a metallicity dependence of the PLRs and simultaneously estimated distances to the GCs adopting different calibrations based on theoretical models and {\it Gaia} EDR3 data. The quantified metallicity dependence of $\sim 0.15-0.17$ mag dex$^{-1}$ is in good agreement with theoretical predictions for the PLZ$_{K_s}$ relation from \citet{marconi2015}. When excluding the calibrator relations, the metallicity dependence of the PLRs in the GCs was found to be sensitive to the adopted spectroscopic metallicities, the RR Lyrae sample size, and the number of clusters, and can vary between $0.07$ and $0.15$~mag dex$^{-1}$. Therefore, homogeneous NIR time-series data in additional GCs, in particular, in metal-rich ([Fe/H]$>-1.0$~dex) and more metal-poor ([Fe/H]$<-2.1$~dex) GCs, is required to extend the metallicity baseline and the statistics of both the number of RR Lyrae and the GCs. 

Our analysis has shown that a robust constraint on the metallicity effects on the PLRs and precise RR Lyrae distances to GCs can be obtained simultaneously, provided a statistically significant sample of GCs covering a wide range of mean-metallicities is available. However, the calibration of the RR Lyrae PLZ relation in the Milky Way is still hampered by uncertainties and a possible zero-point offset in the parallaxes even with {\it Gaia} EDR3. Using theoretical PLZ relations to predict parallaxes for Galactic RR Lyrae, we found a small parallax zero-point offset of $-7\pm3~\mu$as in EDR3 which is significantly smaller than value ($-32\pm4~\mu$as) obtained using the recipe of \citet{lindegren2020a}. We note that our estimated parallax offsets are in agreement with the values provided by \citet{lindegren2020a} corrections for faint and distant RR Lyrae stars that have small parallaxes ($<0.5~$mas). When {\it Gaia} EDR3 parallaxes were corrected using \citet{lindegren2020a} formulae, the median difference between the predicted (i.e. from theoretical PLZ relation) and EDR3 parallaxes increased to $22\pm2~\mu$as, suggesting an overcorrection in the parallaxes. Although the parallaxes have improved significantly in {\it Gaia} EDR3, understanding possible systematics is crucial to obtain the absolute calibration of RR Lyrae PLZ relations. At the same time, homogeneous time-series observations and spectroscopic metallicities of Galactic RR Lyrae are also desired to complement the improved {\it Gaia} parallaxes. Our ongoing program to obtain time-series data in more GCs will help improve the metallicity constraints on the NIR RR Lyrae PLRs in all three NIR filters and simultaneously provide a homogeneous RR Lyrae distance scale to GCs with future {\it Gaia} data releases.

\acknowledgements

We thank Adam Riess, Lennart Lindegren, and Coryn Bailer-Jones for their comments or discussions on the earlier version of this manuscript, which helped in improving the paper. AB acknowledges a Gruber fellowship 2020 grant sponsored by the Gruber Foundation and the International Astronomical Union. AB is supported by the EACOA Fellowship Program under the umbrella of the East Asia Core Observatories Association, which consists of the Academia Sinica Institute of Astronomy and Astrophysics, the National Astronomical Observatory of Japan, the Korea Astronomy and Space Science Institute, and the National Astronomical Observatories of the Chinese Academy of Sciences. HPS and SMK acknowledge support from the Indo--US Science and Technology Forum, New Delhi, India. CCN is grateful for funding from the Ministry of Science and Technology (Taiwan) under contract 109-2112-M-008-014-MY3. This research was supported by the Munich Institute for Astro- and Particle Physics (MIAPP) of the DFG cluster of excellence ``Origin and Structure of the Universe''. This research uses data obtained through the Telescope Access Program (TAP) of China.

This work has made use of data from the European Space Agency (ESA) mission {\it Gaia} (\url{https://www.cosmos.esa.int/gaia}), processed by the {\it Gaia} Data Processing and Analysis Consortium (DPAC, \url{https://www.cosmos.esa.int/web/gaia/dpac/consortium}). Funding for the DPAC has been provided by national institutions, in particular the institutions participating in the {\it Gaia} Multilateral Agreement. 

\facility{CFHT (WIRCam Near-infrared imager)}

\software{\texttt{IRAF} \citep{tody1986, tody1993}, \texttt{DAOPHOT/ALLSTAR} \citep{stetson1987} \texttt{DAOMATCH and DAOMASTER} \citep{stetson1993}, \texttt{ALLFRAME} \citep{stetson1994}, \texttt{SExtractor} \citep{bertin1996}, \texttt{SWARP} \citep{bertin2002}, \texttt{SCAMP} \citep{bertin2006}, \texttt{WeightWatcher} \citep{marmo2008}, \texttt{IDL} \citep{landsman1993}, \texttt{Astropy} \citep{astropy2013}}\\\\\\\\

\vspace{100pt}

\bibliographystyle{aasjournal}
\bibliography{/home/anupam/work/manuscripts/mybib_final.bib}
\end{document}